\documentclass[12pt, preprint]{aastex}
\usepackage{amssymb}
\usepackage{amsmath}
\shorttitle{Adiabatically Contracted Rotation Curves}
\shortauthors{Puglielli}

\begin{document}

\title{Detecting Adiabatic Contraction in Rotation Curves}
\author{David Puglielli\altaffilmark{1}
\affil{Department of Astronomy, University of Cape Town, Rondebosch 7700, South Africa}}

\altaffiltext{1}{david.puglielli@uct.ac.za}
%\altaffiltext{2}{widrow@astro.queensu.ca}

\begin{abstract}

We examine the structure of adiabatically contracted Einasto profiles, using the prescriptions of Blumenthal et al. (1986) and Gnedin et al. (2004), and its impact on rotation curves. Adiabatically contracted halos display a central power index of $\sim0.7\pm0.1$ for nearly all values of the Einasto shape parameter $\alpha$, and are well fit inside $\lesssim1$ kpc by double power laws. However, attempts to fit exponential disc and uncontracted halos to adiabatically contracted rotation curves yield disc masses and central power indices that are too large. We also determine whether or not the rotation curve of NGC 6503 displays evidence of adiabatic contraction using previously published bar formation constraints, and find that NGC 6503 most likely has a minimally contracted halo. However, this conclusion depends on the correct choice of circular velocity curve.

\end{abstract}


\section{Introduction}

One of the most important unsolved problems of cosmology and galaxy evolution is how baryons affect dark matter halos. Over the past two decades, high resolution cosmological simulations have effectively settled the question of what form dark matter halos take in the $\Lambda$CDM cosmogony, which has been very successful in accounting for the large scale structure of the universe. However, on galactic scales, detailed studies of the structure of galaxy halos, as inferred from rotation curves and lensing measurements, have produced mixed results, and it is critical to understand how baryons affect dark matter halos to resolve the discrepancy. 

Traditionally, double power laws have been used to fit cosmological $N-$body halos \citep{navarroetal97, mooreetal, diemandetal}, the most general form for which is given by
\begin{equation}
 \rho(r) = \frac{\rho_{\mathrm{s}}}{\left({r}/{r_{\mathrm{s}}}\right)^{\gamma} \left(1+\left({r}/{r_{\mathrm{s}}}\right)^{\beta}\right)^{\frac{\delta-\gamma}{\beta}}} 
\label{eq:dpow}
\end{equation}
where $\rho_{\mathrm{s}}$ and $r_{\mathrm{s}}$ are the scale density and radius respectively, and the triplet $(\beta,\gamma,\delta)$ governs the shape of the log-log profile.  Commonly, $\beta$ and $\delta$ are set to 1 and 3 respectively, allowing only the central power index to vary:
\begin{equation}
 \rho(r) = \frac{\rho_{\mathrm{s}}}{\left({r}/{r_{\mathrm{s}}}\right)^{\gamma} \left(1+{r}/{r_{\mathrm{s}}}\right)^{3-\gamma}}.
\label{eq:nfw}
\end{equation}
We call Eq.~\ref{eq:nfw} the NFW-type profile. Setting $\gamma=1$ produces the traditional NFW formula \citep{navarroetal97}, and the resulting two parameter profiles were thought to be universal. Recently, high resolution simulations have suggested that simulated halos vary slightly in shape, thus requiring an extra shape parameter \citep{merrittetal, navarroetal10}. The adopted profile is usually the \citet{einasto} profile, given by
\begin{equation}
 \rho(r) = \rho_{\mathrm{h}} \exp \left(-\frac{2}{\alpha}\left[\left(\frac{r}{r_{\mathrm{h}}}\right)^\alpha-1\right]\right).
\label{eq:einasto}
\end{equation}
This form is identical to that used by \citet{navarroetal10}; $r_{\mathrm{h}}$ and $\rho_{\mathrm{h}}$ are the radius and density of the peak of the $r^2\rho$ profile, respectively, and $\alpha$ is the shape parameter. Eq.~\ref{eq:einasto} is formally identical to the classical S\'{e}rsic (1968) $1/n$ profile that is often used to model the projected surface brightnesses of early type galaxies and bulges; however, the S\'{e}rsic profile is applied to the projected radius, while the Einasto profile is applied to the spatial radius. For studies of halo structure, Eq.~\ref{eq:einasto} is preferable because the constants are directly related to the easily calculated $r^2\rho$ quantity.

Modifications to Eq.~\ref{eq:einasto} will occur because of baryon condensation as galaxies form. The simplest prescription for baryon-affected dark matter halos is adiabatic contraction, which posits that halo orbits are purely spherical and do not cross, and that baryon condensation occurs adiabatically, and therefore that angular momentum is conserved \citep{blumenthaletal}. The adiabatic contraction condition may be expressed as
\begin{equation}
  r_{\mathrm{i}} M(r_{\mathrm{i}}) = r_{\mathrm{f}}\left[ M_{\mathrm{f}}(r_{\mathrm{f}}) + M_{\rm{b}}(r_{\mathrm{f}})\right] 
\label{eq:ac}
\end{equation}
where $M(r_{\mathrm{i}})$ is the initial mass of dark matter and baryons at radius $r_{\mathrm{i}}$, $M_{\mathrm{f}}(r_{\mathrm{f}})$ is the final mass of dark matter at radius $r_{\mathrm{f}}$, and $M_{\rm{b}}(r_{\mathrm{f}})$ is the mass of the final baryon configuration at $r_{\mathrm{f}}$ (usually $M_{\rm{b}}$ is an exponential disc, but may in principle be any distribution). The equation may be easily solved for $r_{\mathrm{f}}$ using iterative methods. Adiabatic contraction leads to a pronounced concentration of mass in the centre of a halo, steepening the cusp.

While simple, adiabatic contraction is unlikely to provide a complete description of how baryons impact dark matter halos, because it does not account for radial motions (essentially assuming that the radial action is zero). Additionally, simulations suggest that angular momentum is not, in fact, conserved \citep{gnedinetal, duffyetal, booketal}. Modified adiabatic contraction prescriptions, such as those introduced by \citet{gnedinetal}, and simulations, such as those done by \citet{sellwoodmcgaugh} and \citet{abadietal}, still predict a relative contraction of matter in the centre, but not as much as Blumenthal's original prescription because the radial motions function as a pressure support resisting compression. \citet{gnedinetal} modify the adiabatic contraction prescription by replacing $r$ in Eq.~\ref{eq:ac} with $\bar{r}$, the orbit-averaged radius, as follows:
\begin{equation}
  r_{\mathrm{i}} M(\bar{r}_{\mathrm{i}}) = r_{\mathrm{f}}\left[ M_{\mathrm{f}}(\bar{r}_{\mathrm{f}}) + M_{\rm{b}}(\bar{r}_{\mathrm{f}})\right]
\label{eq:gnedin}
\end{equation}
The relationship between $r$ and $\bar{r}$ is given approximately by $\bar{x}=Ax^w$, where $x\equiv r/r_{\rm{vir}}$, $A=0.8$ and $w=0.8$. This substitution is designed to account for the fact that halo orbits are generally not spherical. \citet{abadietal} find a relationship given by
\begin{equation}
 \frac{r_{\mathrm{f}}}{r_{\mathrm{i}}}=1-B\left[\left(\frac{M_{\mathrm{f}}}{M_{\mathrm{i}}}\right)^c-1\right]
\label{eq:abadi}
\end{equation}
where $B=0.3$, $c=2$ and $M_{\mathrm{f}}/M_{\mathrm{i}}=M(r_{\mathrm{i}})/[M_{\mathrm{f}}(r_{\mathrm{f}})+M_{\rm{b}}(r_{\mathrm{f}})]$; this is, however, a statistical average of several simulations. The \citet{abadietal} prescription produces still less contraction than the \citet{gnedinetal} prescription.

The full effect of baryons on dark matter halos remains an open question, as multiple effects must be accounted for to obtain a full picture of this process. There is ample evidence from $N-$body simulations that the contraction can probably be reversed by supernova feedback in the centres of galaxies \citep{navarroetal96, readgilmore, mashchenkoetal, governatoetal}, while other authors propose that the contraction may be reversed or reduced by infalling gaseous clumps \citep{elzantetal1, elzantetal2, elmegreenetal}, by the effect of a bar torquing the halo \citep{weinbergkatz, holleybockelmannetal}, by preprocessing of primordial halo clumps \citep{momao}, or by the combined effect of dynamical friction and angular momentum \citep{delpopolokroupa}. These factors complicate attempts to use adiabatic contraction and related theories to model observed rotation curves. Indeed, most measurements of galaxy halo profiles conclude that LSB galaxies have much more shallow cusps than implied by cosmological simulations \citep{floresprimack, debloketal01, simonetal, kuziodenarayetal}, suggesting that adiabatic contraction is reversed; note that adiabatic contraction \emph{always} produces cuspy halos, even if the uncontracted haloes are cored (cf. Fig.~2 of Blumenthal et al. 1986 and Dutton 2005). However, \citet{ohetal} find that they can reproduce the observations of LSB galaxies by including star formation, supernova feedback and extragalactic UV heating, all of which overwhelm the effect of adiabatic contraction. Meanwhile, several studies of HSB galaxies suggest that  the data is consistent with cuspy and cored profiles, even with multiple constraints factored in \citep{duttonetal1, debloketal, augeretal, schulzetal}. 

There is moreover no definitive proof on whether adiabatic contraction is a valid general description of baryon affected dark matter halos. For disc galaxies, \citet{kassinetal} find that adiabatic contraction yields poorer fits to the rotation curve, especially if the disc dominates, while \citet{duttonetal1} find rotation curves with submaximal discs equally well fit by both contracted and uncontracted halos. \citet{schulzetal} find that adiabatically contracted NFW profiles provide an excellent fit for their sample elliptical galaxies, while \citet{augeretal} find that adiabatic contraction significantly overestimates the dark matter content of early type galaxies. The discrepant conclusions reflect a broader degeneracy between halo contraction and the choice of IMF \citep{napolitanoetal, duttonetal2}, but the very real possibility exists that the effects of baryons change significantly between different classes of galaxies, and even between different galaxies of the same class: the simulations of \citet{abadietal} and \citet{tisseraetal} suggest that the halo assembly history, which is certainly not identical for all galaxies, plays a crucial role in determining the halo response to the baryons. %However, a striking feature of these studies is the considerable variation in parameters derived for different galaxies, suggesting that there are significant variations in quantities such as the baryon fraction in primordial halos. Thus, the details of baryon condensation in any particular galaxy may depend on environmental factors.

In this paper, we restrict our analysis to how Einasto profiles and their associated rotation curves are affected by the adiabatic contraction prescriptions of Blumenthal et al. and Gnedin et al. In particular, we would like to know if it is possible to unambiguously identify adiabatically contracted rotation curves in disc galaxies, or if sufficient uncertainty exists that this is not possible. The structure of this paper is as follows. In Section 2, we discuss how density profiles and rotation curves are affected by adiabatic contraction, using a model halo to investigate these issues. In Section 3, we use published rotation curve data for NGC 6503 to determine if the dark matter content can be constrained. Section 4 discusses and summarizes our conclusions.

\section{The Structure of Adiabatically Contracted Halos and Rotation Curves}

There are three parameters characterizing adiabatically contracted (AC) halos: the baryon fraction $f_{\mathrm{b}}$, the baryon scale radius $r_{\mathrm{b}}$ (which is the exponential scale length in our case), and the radius out to which baryons condense, $r_{\rm{outer}}$. This last radius need not in principle be the virial radius $r_{\rm{vir}}$.\footnote{We define the virial radius as the radius within which the mean density of the halo is 200 times the critical density $\rho_{\rm{crit}}$.} To examine the structure of AC halos, we begin with a model Einasto profile (Eq.~\ref{eq:einasto}) with $r_{\mathrm{h}}=22.65$, $\rho_{\mathrm{h}}=0.00058265$ and $\alpha=0.17$. These parameters are chosen to match the mass-concentration relation of \citet{maccioetal} and to produce a halo and disc combination whose rotation curve peaks at $\sim$220 km s$^{-1}$. For comparison, we include a halo with identical $\rho_{\rm{h}}$ and $r_{\rm{h}}$ with $\alpha=0.5$. We then apply the Blumenthal and Gnedin adiabatic contraction prescriptions at varying degrees of contraction given by $f_{\mathrm{b}}^1=0.01$, $f_{\mathrm{b}}^2=0.075$, and $f_{\mathrm{b}}^3=0.15$. We use GalactICS units for this exercise \citep{widrowetal}, in which $G=1$, the unit of length is 1 kpc and the unit of velocity is 100 km~s$^{-1}$, so that the unit of mass is equal to $2.325\times10^9M_{\odot}$.

\begin{figure}
\begin{center}
\includegraphics[angle=270,scale=0.6]{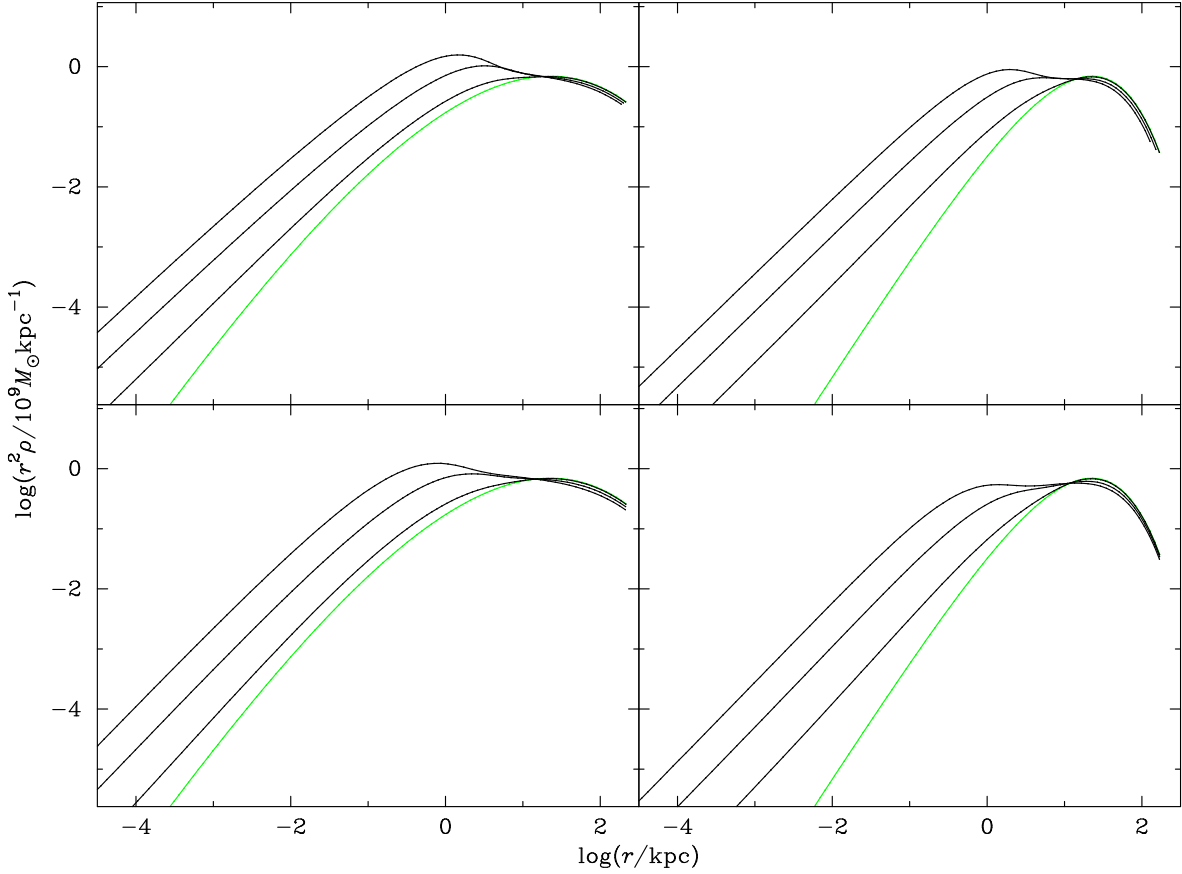}\end{center}
\caption{Examples of adiabatic contraction applied to two different Einasto profiles, at differing degrees of contraction. The top panels use Blumenthal's prescription, while the bottom panels use Gnedin's prescription. For both halos, $\rho_{\mathrm{h}}=0.00058265$, $r_{\mathrm{h}}=22.65$; for the halo on the left, $\alpha=0.17$, while the right panels use a halo for which $\alpha=0.5$. The contracted halos are given by $r_{\mathrm{b}}=3$, $f_{\mathrm{b}}^1=0.01$, by $r_{\mathrm{b}}=3$, $f_{\mathrm{b}}^2=0.075$, and by $r_{\mathrm{b}}=1$, $f_{\mathrm{b}}^3=0.15$, in order from smallest to largest degree of contraction, for all four panels. Green identifies the original halos and black identifies the contracted halos. The contracted halos display a relatively constant inner power and, in the case of stronger contraction, a double inflection near or just outside the peak. Einasto profiles cannot reproduce either of these features. }
 \label{fig:acprofile}
 \end{figure}

Fig.~\ref{fig:acprofile} shows how the $r^2\rho$ profile of these AC halos compares to standard Einasto profiles. % at different values of $\alpha$ and NFW-type profiles (Eq.~\ref{eq:nfw}) at different values of $\gamma$, all scaled to the same peak. It is clear from this figure that neither the standard Einasto profile nor the NFW-type profile can reproduce AC profiles, because 
At higher degrees of contraction, AC profiles display a double inflection just beyond the peak of the $r^2\rho$ profile that neither Eq.~\ref{eq:dpow} nor Eq.~\ref{eq:einasto} can reproduce. The double inflection is not apparent on a $\log\rho-\log r$ plot, but is very clear on a $\log r^2\rho$ plot. Adiabatically contracted profiles also display a nearly linear structure inside the $r^2\rho$ peak, indicating power law behaviour. Einasto profiles cannot reproduce the power law behaviour, but NFW-type and double power laws exhibit power law behaviour as $r\rightarrow0$. %A similar power law/double inflection structure is observed for the modified adiabatic contraction prescription of \citet{gnedinetal}.%\footnote{The Einasto profile itself asymptotically approaches a power law as $r\rightarrow 0$, but the power law form of the AC profiles is apparent well outside the radius of convergence of cosmological simulations.} 

\begin{figure}
\begin{center}
 \includegraphics[angle=270,scale=0.6]{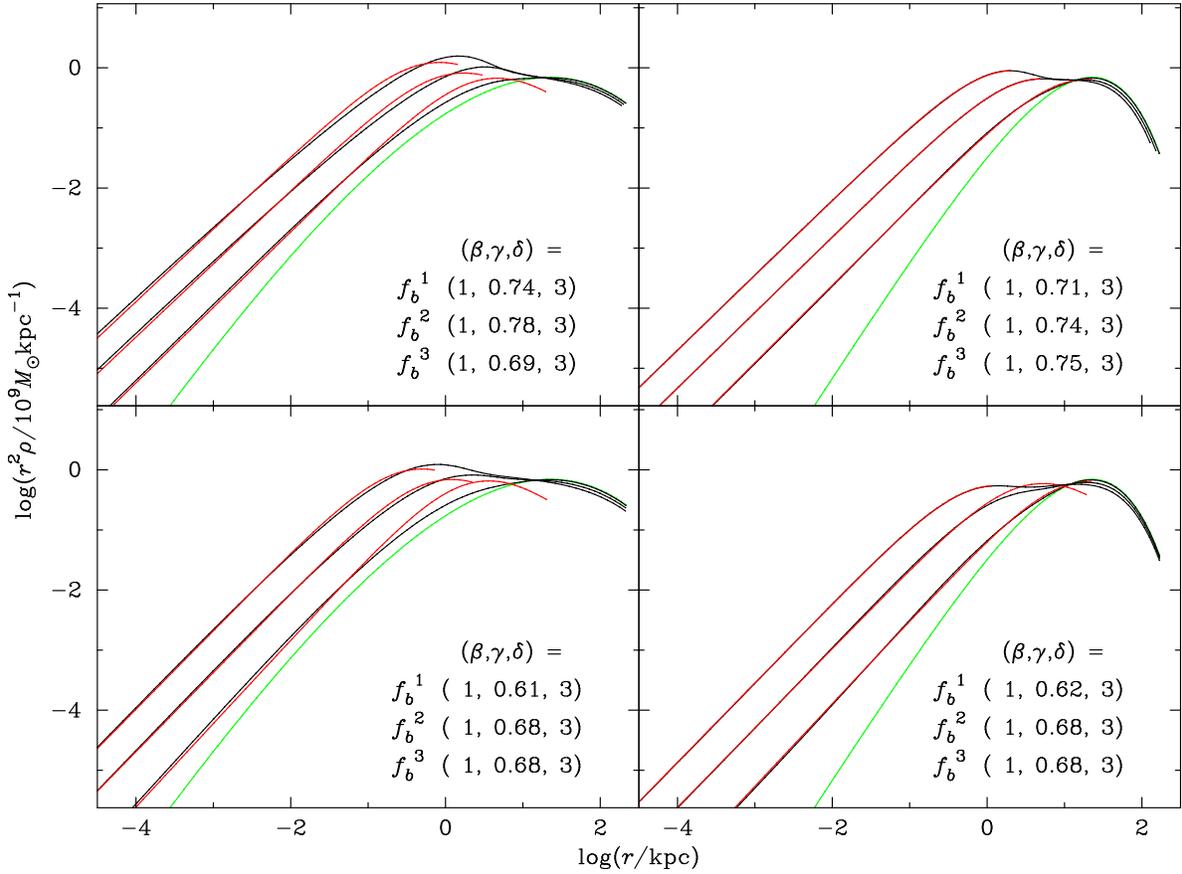}\end{center}
 \caption{Fits to two different Einasto profiles at differing degrees of contraction, using NFW-type profiles (Eq.~\ref{eq:nfw}). The top panels use Blumenthal's prescription, while the bottom panels use Gnedin's prescription. The profiles and colours are as found in Fig.~\ref{fig:acprofile}. Red identifies the fits. The best fit values for $\gamma$ are provided for each contracted halo. The fits are generally fairly poor, although they improve at higher degrees of contraction and for higher $\alpha$.}
 \label{fig:acnfwfits}
 \end{figure}

\begin{figure}
\begin{center}
 \includegraphics[angle=270,scale=0.6]{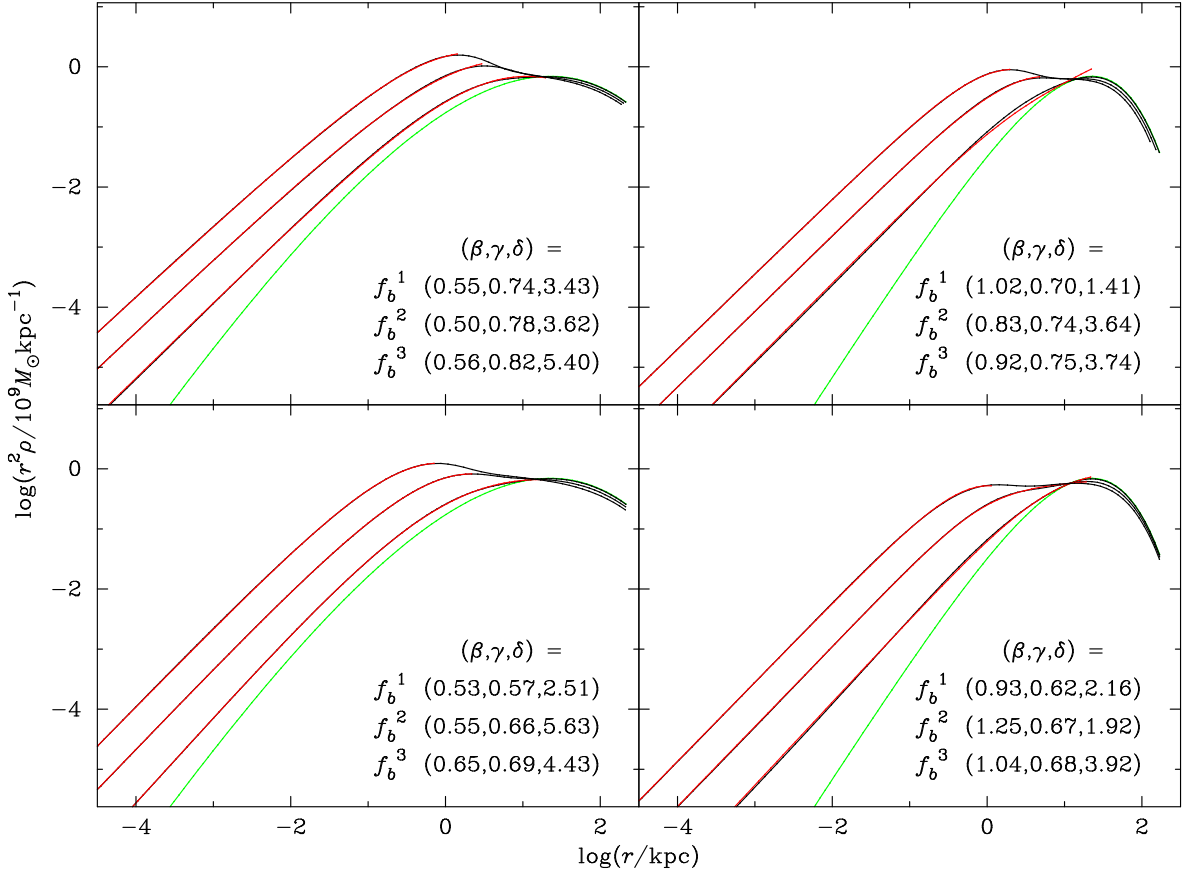}\end{center}
 \caption{Fits to two different Einasto profiles at differing degrees of contraction, using double power laws (Eq.~\ref{eq:dpow}). The profiles and colours are as found in Fig.~\ref{fig:acprofile}. Red identifies the fits. The best fit values for $\beta$, $\gamma$ and $\delta$ are provided for each contracted halo. The fits are considerably better than those found in Fig.~\ref{fig:acnfwfits}; they tend to break down at higher $\alpha$ and at smaller degreees of contraction.}
 \label{fig:acdpowfits}
 \end{figure}

%Therefore, inside the radius of the peak of the $r^2\rho$ profile, the density of an AC halo can be written as $\rho\propto r^{-\gamma}$ as $r\rightarrow0$. 
To determine if NFW-type or double power laws can fit the inner part of an AC halo, we plot fits to AC halos in Fig.~\ref{fig:acnfwfits} using Eq.~\ref{eq:nfw} and in Fig~\ref{fig:acdpowfits} using Eq.~\ref{eq:dpow}. For both prescriptions, the NFW-type profile fails to provide an effective fit, but the double power law yields a much better fit, provided $\beta\sim0.5-1$ for $\alpha=0.17$ and $\beta\sim1.5-2$ for $\alpha=0.5$. The fits become inferior as $f_{\mathrm{b}}$ decreases (that is, in cases where the final halo is close to the original) at larger $\alpha$, but are otherwise very good. Thus, the central power $\gamma_{\rm{c}}$ obtained by direct calculation is identical to $\gamma$ in Eq.~\ref{eq:dpow}.

\begin{figure}
\begin{center}
 \includegraphics[angle=270,scale=0.6]{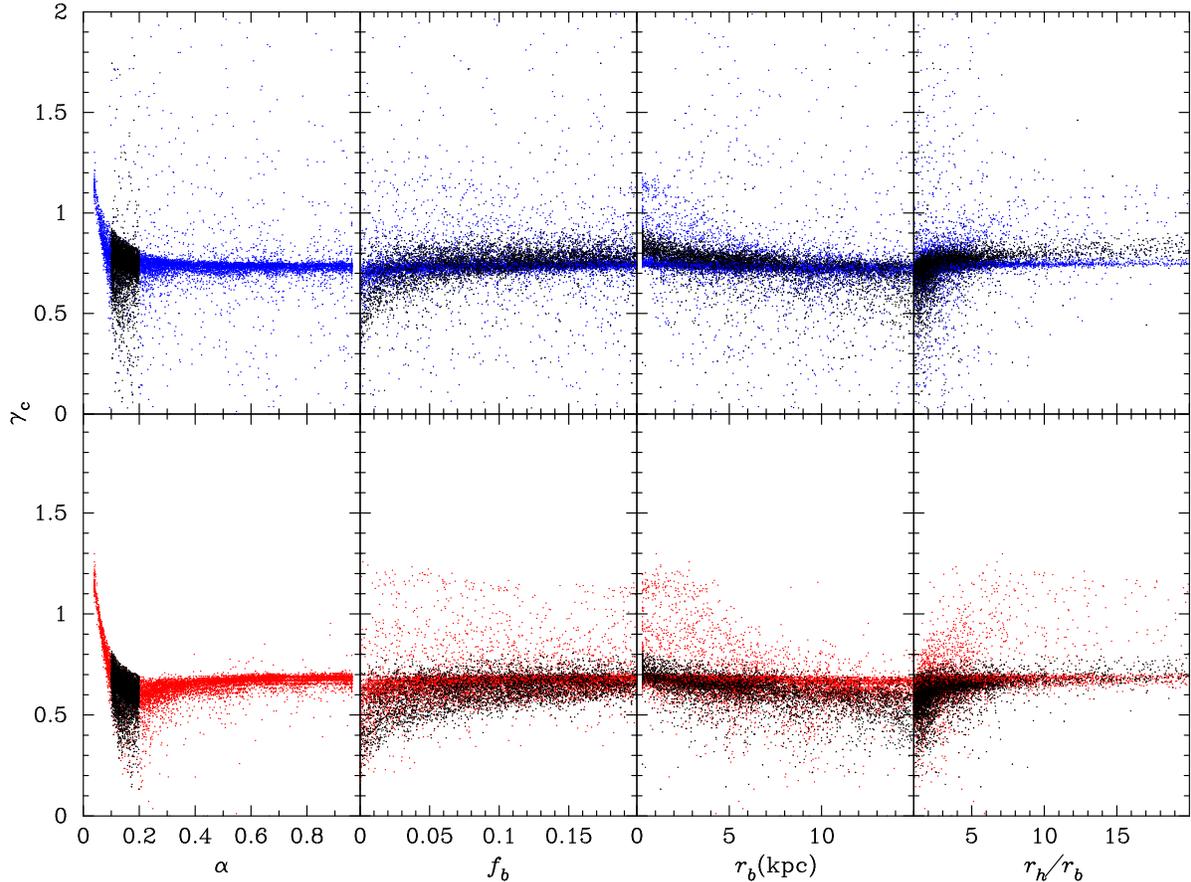}\end{center}
 \caption{Scatter plot of the central power, $\gamma_c$, as a function of the shape parameter $\alpha$, baryon fraction $f_{\mathrm{b}}$, baryon scale length $r_{\mathrm{b}}$ and baryon to halo scale length ratio $r_{\mathrm{b}}/r_{\mathrm{h}}$ from left to right. The top panels refer to Blumenthal's AC prescription, while the bottom panels refer to Gnedin's AC prescription. The black points on all panels identify models for which $0.1<\alpha<0.2$, which is consistent with cosmological simulations. The values for $\gamma_c$ are largely constant with each of these quantities, and the aggregate values are given by $\gamma_c=0.75\pm0.17$ and $0.66\pm0.08$ for the Blumenthal and Gnedin prescriptions, respectively. Cases where $r_{\mathrm{b}} > r_{\mathrm{h}}$ and $r_{\mathrm{b}} > r_{\rm{outer}}$ are excluded from the plot.}
 \label{fig:gammacfour}
 \end{figure}

The values taken by $\gamma$ display a remarkable consistency. Fig.~\ref{fig:gammacfour} provides a scatter plot showing how the central power $\gamma_{\rm{c}}$ varies with $\alpha$, $f_{\mathrm{b}}$, $r_{\mathrm{b}}$ and $r_{\mathrm{b}}/r_{\mathrm{h}}$. At $\alpha\gtrsim0.3$, $\gamma\sim0.8$ as $r\rightarrow0$. For $\alpha\lesssim0.3$, $\gamma$ displays larger variations and becomes larger as $\alpha\rightarrow0$. The consistency of $\gamma$ is maintained across the other three parameters in the figure, including cases for which $0.1<\alpha<0.2$, suggesting that $\gamma$ is not strongly dependent on any of these parameters. %Fig.~\ref{fig:gammavarfour} shows how $\gamma$ varies with virial radius, as a function of the same parameters as in Fig.~\ref{fig:gammacfour}. These variations are expressed as the ratio $\gamma_i/\gamma_0$, where $\gamma_i$ is the power at $10^{-i}r_{\rm{outer}}$, and $\gamma_c$ is the power as $r\rightarrow0$. The AC profiles generally do not display a constant power, varying by up to a factor of $2-3$ at $10^{-2}R_{\rm{outer}}$. 
% at or outside the radius of convergence. Fig.~\ref{fig:varsloperad} shows how $\gamma$ varies with $10^{-3}r_{\rm{outer}}$ and $10^{-2}r_{\rm{outer}}$; the variation is calculated as the ratio $\gamma_i/\gamma_0$, where $\gamma_i$ is the power at $10^{-i}r_{\rm{outer}}$, and $\gamma_0$ is the power as $r\rightarrow0$. There is considerable scatter on both plots; some points lie closer to $\gamma=0$, and these points correspond to cases where $\alpha$ is close to 1 and $r_{\mathrm{b}}/\rho_{\mathrm{h}} \gtrsim 1$, that is, cases where adiabatic \emph{expansion} may occur. However, most points lie on a line that deviates slightly from $\gamma_{-3}/\gamma_0 = 1$.

The consistency of $\gamma$ suggests a possible test for adiabatic contraction. %Because $\gamma$ varies by factors of a few over a narrow range of radius (Fig.~\ref{fig:gammavarfour}), direct measurements of the power are infeasible. However, Fig.~\ref{fig:vanillafits} shows that, inside the peak of the $r^2\rho$ profile, AC profiles may be well fit by double power laws. In this case, $\gamma_c$ on Fig.~\ref{fig:gammacfour} is identical to $\gamma$ in Eqs.~\ref{eq:nfw} and \ref{eq:dpow}. Fig.~\ref{fig:vanillafits} shows double power law fits for a halos of $\alpha=0.17$ and $\alpha=0.5$ at various degrees of contraction: $f_{\mathrm{b}}=0.01, 0.075$ and 0.15. The fits are generally very good, provided $\beta\sim0.5-1$. The fits become inferior as $f_{\mathrm{b}}$ decreases (that is, in cases where the final halo is close to the original), especially at larger $\alpha$. Fig.~\ref{fig:aconefit} shows the best fit for the AC model shown in Fig.~\ref{fig:acprofile}, along with fits to the Gnedin-contracted halo; the best fit corresponds to $\beta=0.55$, $\gamma=0.80$ and $\delta=3.2$. Fits using Eq.~\ref{eq:nfw} are noticeably inferior, but they do reproduce the central power. 
Double power law fits to the rotation curve inside the peak of the $r^2\rho$ profile could be used to determine $\gamma$ and thus if a given disc galaxy displays evidence of adiabatic contraction. The average $\gamma$ for the Blumenthal and Gnedin prescriptions is $0.75\pm0.17$ and $0.66\pm0.08$, respectively. For cases where $0.1<\alpha<0.2$, the Blumenthal and Gnedin prescriptions yield $\gamma=0.75\pm0.17$ and $0.61\pm0.08$ respectively. Although there is considerable overlap in the two values for $\gamma$, Blumenthal's prescription yields larger $\gamma$ with more scatter, suggesting that it may be possible to distinguish between the two types of adiabatic contraction prescriptions as well. Using Eq.~\ref{eq:dpow} rather than Eq.~\ref{eq:nfw} is critical, however, because on observationally accessible scales $\gamma$ is not constant, and because Eq.~\ref{eq:nfw} can only provide an effective $\gamma$ over the range of radii probed, which limits its usefulness as a diagnostic parameter. Eq.~\ref{eq:dpow} yields the true $\gamma$ because the variation in power over a given range of radii can instead be captured by the $\beta$ parameter.%Measured at a given fraction of the virial radius, $\gamma$ can be constrained reasonably well to within $\sim30\%$, and the dependence of $\gamma$ with radius . This is shown in Fig.~\ref{fig:sloperadius}. Because the fraction of the virial radius is not easily measured, in Fig.~\ref{fig:sloperb} we also show $\gamma$ as a function of the exponential scale length $r_{\mathrm{b}}$, which is easily obtained. This figure shows a dependence of $\gamma$ on $r_{\mathrm{b}}$ approximating an exponential decay; $\gamma_{-3}$ is almost linear. Although a trend is clear, there is a great deal of scatter in these plots. Therefore, we also present the same plots restricted to cases where $0.1<\alpha<0.2$, which is consistent with cosmological simulations, in Fig.~\ref{fig:sloperadius-trunc} and \ref{fig:sloperb-trunc}. We also display the same quantities for uncontracted halos in Figs.~\ref{fig:sloperadius-plain} and \ref{fig:sloperb-plain}, showing how these quantities occupy a larger region of this parameter space. Thus, the central power is significantly constrained by both Blumenthal and Gnedin AC prescription. 

\begin{figure}
\begin{center}
 \includegraphics[angle=270,scale=0.6]{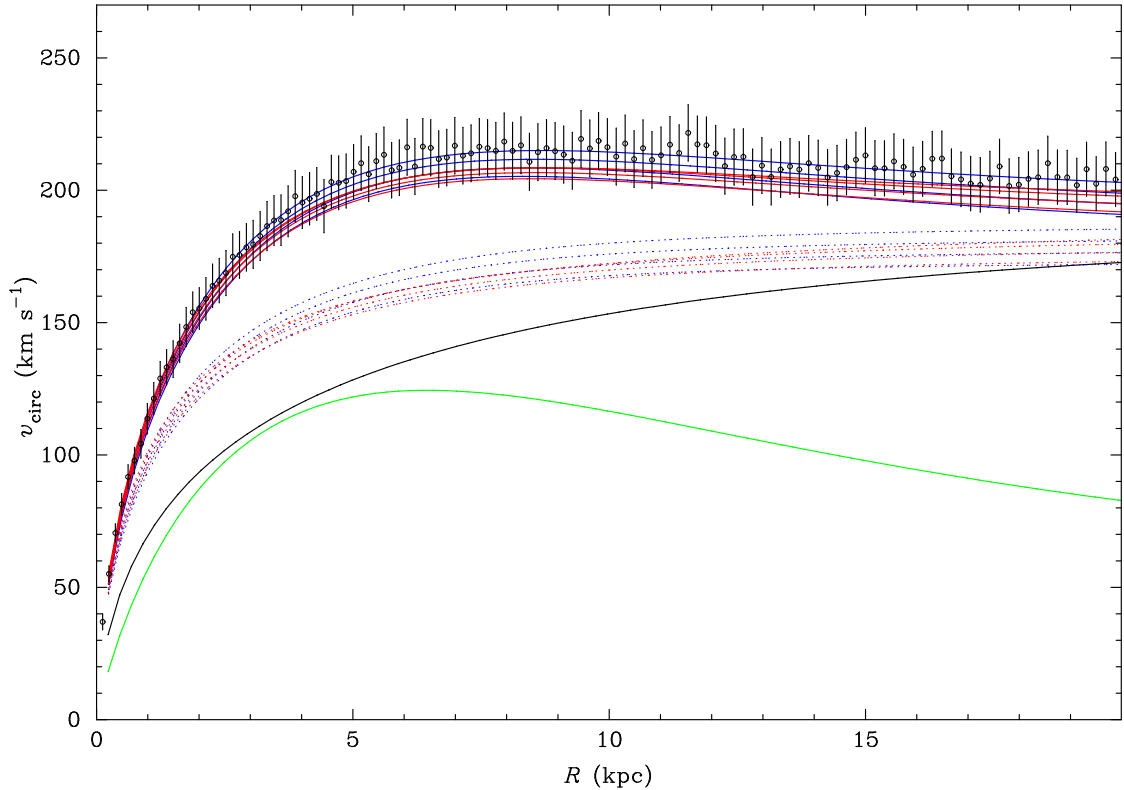}\end{center}
 \caption{The original halo rotation curve (black), contracted halo rotation curves (dotted red and blue) and the full (solid red and blue) AC rotation curves for the Blumenthal prescription (blue) and Gnedin prescription (red), at $f_{\mathrm{b}}=0.025$, 0.05, 0.075, and 0.1 from top to bottom. Noise is added to these curves to create realistic mock data, but this is only shown for the highest rotation curve for clarity. The disc rotation curve for each model is identical and shown in green.}
 \label{fig:testrotation}
 \end{figure}

However, the main probe of halo structure is rotation curves, which depend on the total mass (i.e., the integral of $\rho$) and hence the double inflection, the subtle differences between NFW-type and double power laws, and the differences between the Blumenthal and Gnedin prescriptions may be difficult to detect in real galaxies. %We plot rotation curves of Einasto halos and a model AC halo in Fig.~\ref{fig:acrotation}. The plot suggests that the rotation curves of AC halos do not match the rotation curves of pure Einasto or NFW-type profiles. 
Fig.~\ref{fig:testrotation} displays model rotation curves for the Einasto profile found in the left panel of Fig.~\ref{fig:acprofile}, adiabatically contracted at different values of $f_{\mathrm{b}}$. To to further examine this issue, we now attempt to fit pure Einasto and NFW-type profiles to a mock rotation curve for an AC halo, and apply it to a real galaxy in the section following. %We first wish to discern if it is possible to fit AC halos with model Einasto profiles, and then to determine whether it is possible to fit pure Einasto profiles with AC model halos.

\subsection{Rotation Curve Fits to AC Halos}

There have been relatively few attempts in the literature to directly fit Einasto profiles to rotation curves -- \citet{cheminetal} have done so for M31, but they were not able to constrain the halo profile. Moreover, many attempts to fit any kind of density profile to rotation curves ignore the effects of baryon condensation. In the case of LSB galaxies, this approach is probably justified because of the small disc mass. There have been far more attempts to directly fit NFW-type profiles to rotation curves. It is however interesting to determine how well AC halos can be fit by pure Einasto profiles and double power laws.

There are three parameters governing the Einasto profile, and three governing adiabatic contraction as indicated in \S2.1. These six parameters are summarized in Table~\ref{table:parametertable}. It should be noted that $r_{\rm{outer}}$ is not determined by the observed extent of the rotation curve;  $r_{\rm{outer}}$ impacts the total mass of baryons, and therefore differences in this parameter will impact the \emph{entire} rotation curve, even if all other parameters are kept constant.

We adopt the same Einasto halo as in \S2.1 for consistency with the concentration-mass relation, and include a disc of a fixed mass $m_{\rm{disc}}=12$ and scale length $r_{\mathrm{b}}=3$. The resulting rotation curve peaks at about 220 km~s$^{-1}$. We then apply the adiabatic contraction prescription, using the same halo parameters, at different values of $f_{\mathrm{b}}$: $f_{\mathrm{b}} = 0.025$, 0.05, 0.075 and 0.1. We reproduce the initial disc mass by adjusting $r_{\rm{outer}}$, so that the disc mass is identical for all $f_{\mathrm{b}}$. The resulting model rotation curves are sampled linearly in radius with error bars and resolution roughly consistent with THINGS data published in \citet{debloketal}, to which noise is added. We then attempt to fit these rotation curves using pure Einasto profiles, holding the disc scale length fixed since this quantity may be constrained by surface brightness profiles. We use a Bayesian/MCMC approach to find the best fit for each case, and then carry out the same procedure for Gnedin's adiabatic contraction prescription. The original and AC rotation curves are shown in Fig.~\ref{fig:testrotation} for reference.

\begin{figure}
\begin{center}
 \includegraphics[angle=270,scale=0.6]{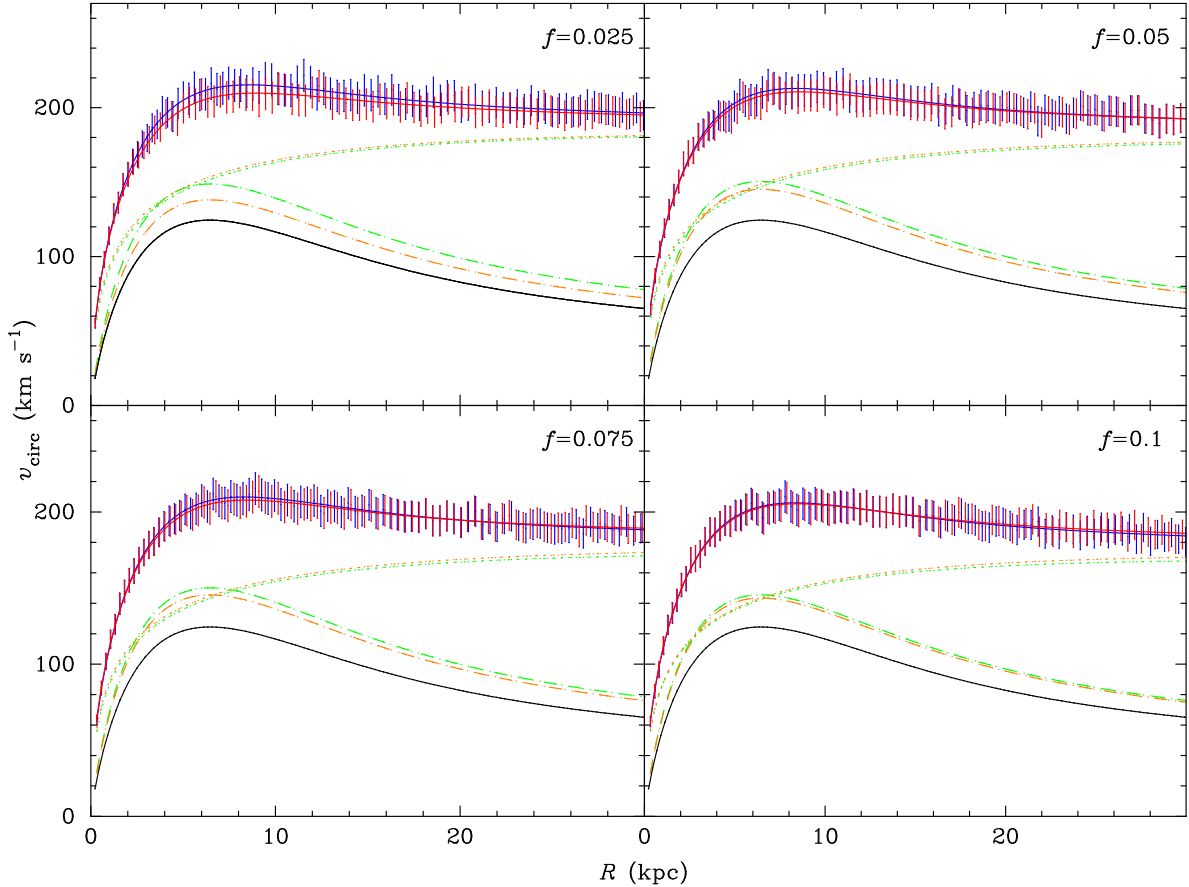}\end{center}
 \caption{Fits to model rotation curve, adiabatically contracted for different values of $f_{\mathrm{b}}$, using plain Einasto profiles. Blue points with error bars identify some of the Blumenthal-contracted mock data points and the blue curve identifies the best fit total rotation curve to the \citet{blumenthaletal} prescription, while red points with error bars identify some of the Gnedin-contracted mock data points and the red curve identifies the best fit total rotation curve to the \citet{gnedinetal} prescription. For clarity, not all the mock data is shown. Green and orange identify the disc and halo breakdown of the rotation curves for the Blumenthal and Gnedin prescriptions, respectively. The disc rotation curve is shown as dashed lines, and the halo rotation curve is shown as dotted lines. The original disc (identical for all cases) is shown as the lowest black curve. The fits are very good, but overestimate the disc mass by $20-40\%$.}
 \label{fig:testdatafits}
 \end{figure}

The resulting fits are found in Fig.~\ref{fig:testdatafits} and the posterior mean and error bars are found in Table~\ref{table:testresults}, with GalactICS masses converted back to physical masses. The fits are excellent, suggesting that AC halos are well fit by Einasto profiles. However, the fits produce smaller $\alpha$ than found in the original halo, and the disc mass required to achieve these fits is about 40\% larger than used to construct the rotation curves in each case, suggesting that attempts to fit AC rotation curves with Einasto profiles overestimate the disc mass. This is because AC rotation curves have steeper slopes near the centre, and heavier discs are required to reproduce this feature when using a pure Einasto halo. Using Gnedin's adiabatic contraction prescription alleviates the problem somewhat, yielding marginally lighter discs. 

While the best fits overestimate the disc mass, it may be possible to obtain good fits by fixing the disc mass to its correct value.\footnote{Note that the error bars in Table~\ref{table:testresults} are formal 1$-\sigma$ error bars obtained from MCMC. Qualitatively, such error bars may not be large enough to provide an acceptable range of values.} %We show fits made using this assumption in Fig.~\ref{fig:testdatafits-fixdisc}. 
In this case, we find that the fits are not as good but improve as $f_{\mathrm{b}}$ increases and improve when Gnedin's prescription is used as well; in general, however, these fits are inferior to those that allow the disc mass to vary and are not consistent across various values of the adiabatic contraction parameters.

%Also shown in Table~\ref{table:testresults} are the $\langle X\rangle$ values of the discs, the Toomre $X$ global stability parameter averaged over two disc scale lengths. The $X$ parameter is a measure of stability to global modes, and provides an approximate measure of whether or not a disc is likely to be bar unstable. Values of $\sim3$ operate as an approximate cutoff for bar formation. The original model has $\langle X\rangle=3.2$, while the fitted models have $\langle X\rangle\sim2.5-3$. Overestimates of the disc mass may imply that discs are bar unstable in galaxies not known to possess a bar; however, as Figs.~\ref{fig:testdatafits}-\ref{fig:testdatafitsdpow} reveal, the rotation curve breakdown remains halo-dominated.

\subsection{Fits Employing NFW-type and Double Power Profiles}

\begin{figure}
\begin{center}
 \includegraphics[angle=270,scale=0.6]{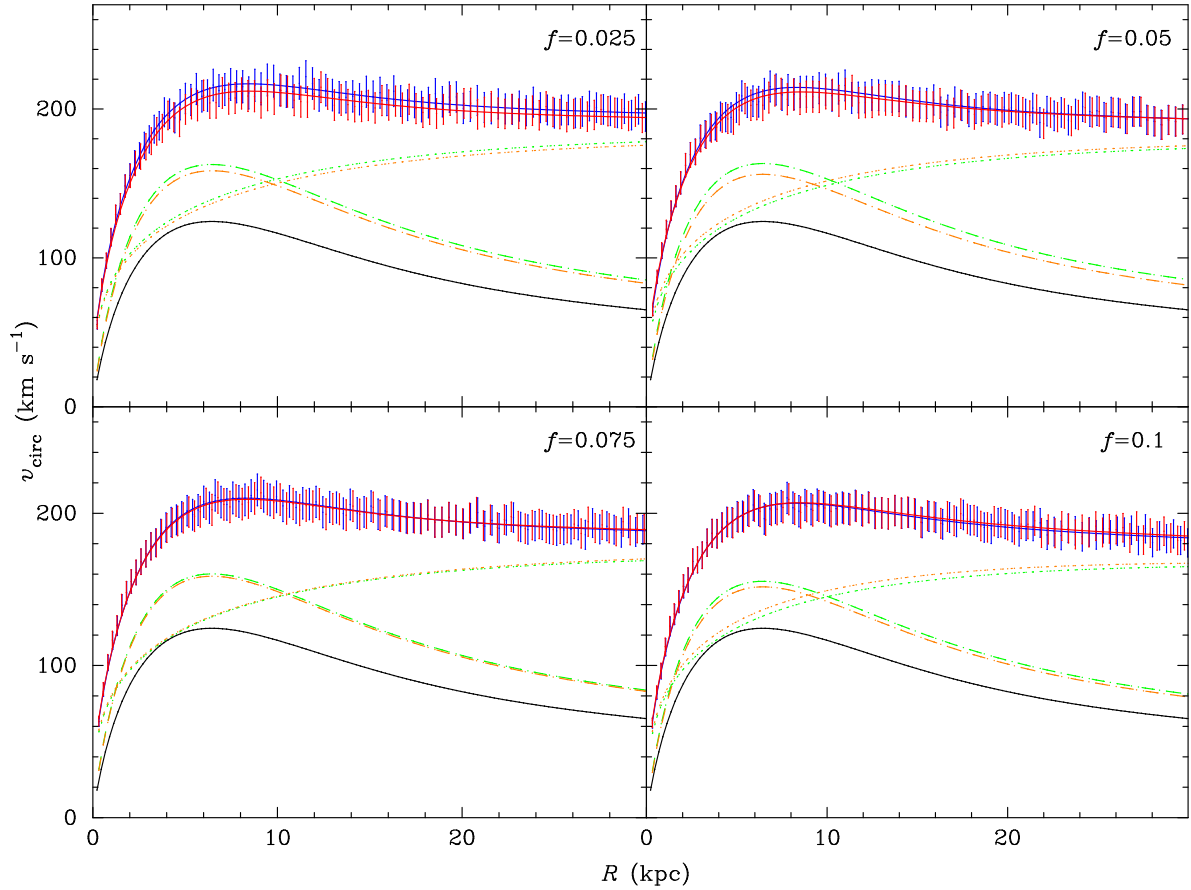}\end{center}
 \caption{Fits to model rotation curve, adiabatically contracted for different values of $f_{\mathrm{b}}$, using NFW-type profiles in which $\beta$ and $\delta$ are held fixed. Colours and line types are as found in Fig.~\ref{fig:testdatafits}. The fits are similar to those in Fig.~\ref{fig:testdatafits}.}
 \label{fig:testdatafitsnfw}
 \end{figure}

\begin{figure}
\begin{center}
 \includegraphics[angle=270,scale=0.6]{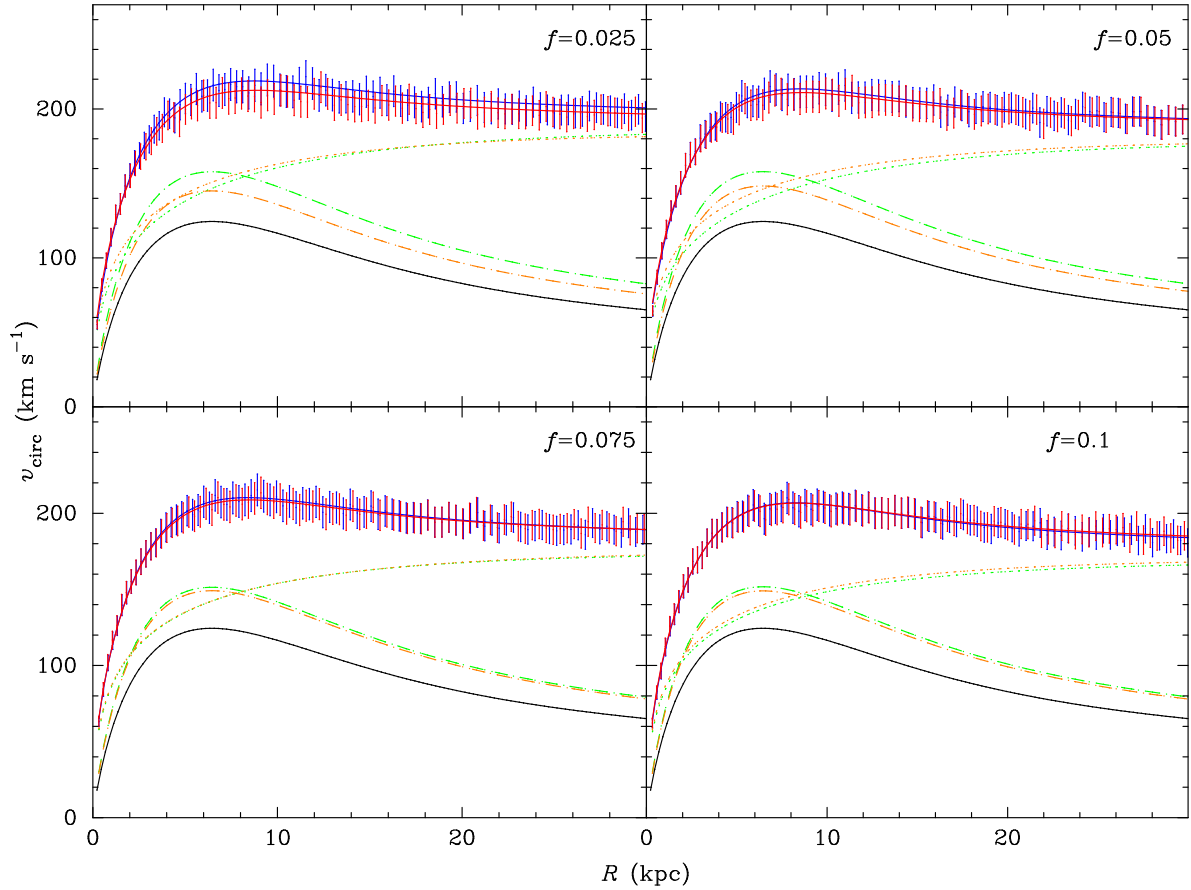}\end{center}
 \caption{Fits to model rotation curve, adiabatically contracted for different values of $f_{\mathrm{b}}$, using double power laws in which all three indices $\beta, \gamma$ and $\delta$ are allowed to vary. Colours and line types are as found in Fig.~\ref{fig:testdatafits}. The fits are similar to those in Figs.~\ref{fig:testdatafits} and \ref{fig:testdatafitsnfw}.}
 \label{fig:testdatafitsdpow}
 \end{figure}

The fact that AC Einasto profiles exhibit power law behaviour inside $r_{\mathrm{h}}$ suggests that Eqs.~\ref{eq:dpow} and \ref{eq:nfw} may produce better fits to the rotation curves of AC halos. We first carry out fits with Eq.~\ref{eq:nfw} in which $\beta$ and $\delta$ are fixed to 1 and 3, while the central power index $\gamma$ is allowed to vary. These fits are otherwise identical to the fits carried out in the previous subsection. The resulting fits are found in Fig.~\ref{fig:testdatafitsnfw} and are essentially identical to the Einasto fits. However, they too overestimate the disc mass, in some cases by more than 50\%. Allowing $\beta$ and $\delta$ to vary still overestimates the disc mass (Fig.~\ref{fig:testdatafitsdpow}). We conclude that NFW-type and double power profiles offer similarly good fits to the rotation curves of AC Einasto halos, although disc masses are overestimated.

Note that $\gamma$ is found to be $\sim$1.2, which exceeds the value of $\sim0.6-0.8$ for these halos, as indicated in Figs.~\ref{fig:acnfwfits} and \ref{fig:acdpowfits}. We therefore also attempt to fit a double power profile with $\gamma$ fixed to 0.8, finding that the fits require very massive discs (more than twice the true mass). Moreover, attempts to fit double power laws to these rotation curves with a fixed disc of the correct mass fail to yield the correct value of $\gamma$. We conclude that although AC profiles may be described by a power law inside a certain radius, NFW-type and double power profiles cannot be used to fit the full rotation curves of AC halos because these fits overestimate the disc mass, and because the fitted $\gamma$ exceeds the true $\gamma$.

\subsection{Fitting only the Inner Rotation Curve}

While using the full rotation curve yields incorrect values for $\gamma$, fits that only use the inner rotation curve may produce better values. Moreover, we would like to determine if it is possible to distinguish between the Blumenthal and Gnedin prescriptions using rotation curve data. Therefore, we carried out the same fitting procedure as in the previous section for double power fits, this time truncating the rotation curve at 5 kpc (well inside the peak of the $r^2\rho$ profile for all halos). 

The resulting values for $\beta, \gamma$, and $\delta$ are found in Table~\ref{table:innerfits}. These may be compared with the `true' values found in Table~\ref{table:truegamma}, which were obtained by fitting the $r^2\rho$ profile directly. We find that $\gamma\lesssim0.9$ for all fits; in most cases, the true $\gamma$ values are not reproduced by the fits, although these estimates are much better than those found in the previous section. However, the error bars are quite large (especially for $\beta$) and the correct values are generally captured at the $2-\sigma$ level. In addition, there is considerable overlap between the fitted values for the Blumenthal and Gnedin prescriptions. Thus, not surprisingly, the mock data cannot distinguish between the two prescriptions, probably because the error bars are not small enough. We conclude that it is essentially impossible to distinguish between the Blumenthal and Gnedin adiabatic contraction prescriptions unless the observational error bars, and possibly resolution, improve substantially.

\section{NGC 6503 as a Test Candidate}

NGC 6503 is an attractive candidate to test how baryons affect dark matter halos. It is mostly isolated, so the impact of mergers and tidal interactions is limited, and therefore a direct comparison of the simple adiabatic contraction prescription is more applicable to this galaxy than other galaxies. Published model circular velocity curves exist from \citet{pugliellietal} (hereafter PWC), which will allow a more accurate comparison of adiabatic contraction with the true mass distribution of the halo. Moreover, the fact that NGC 6503 is clearly strongly cusped invites a direct test of the adiabatic contraction hypothesis.

\begin{figure}
\begin{center}
 \includegraphics[angle=270,scale=0.6]{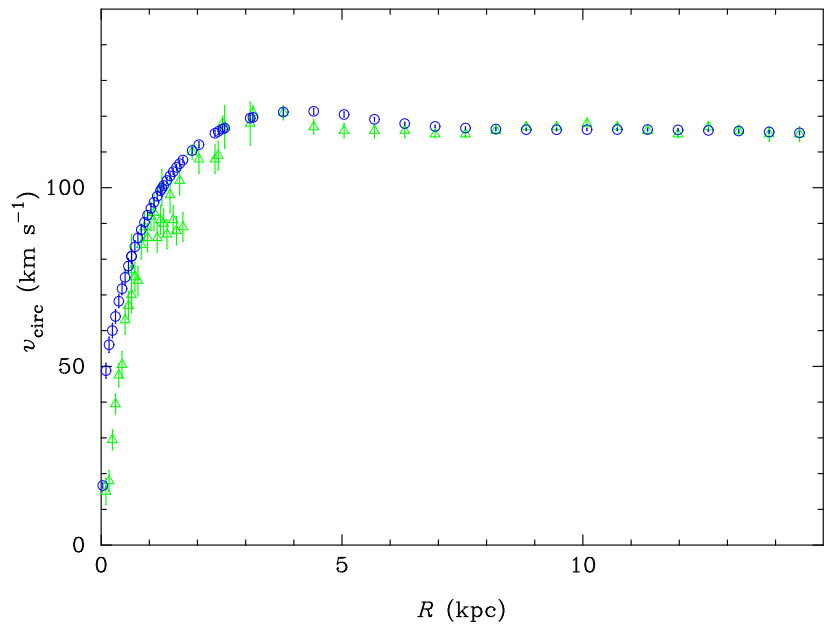}\end{center}
 \caption{The rotation curve for NGC 6503, with the circular velocity from PWC identified by blue circles and the ionized gas curve from \citet{bottema89} and \citet{begeman87} identified by green triangles.}
 \label{fig:6503rotation}
 \end{figure}

The appropriate choice of rotation data must be made. Ideally, we would use data that perfectly traces the gravitational potential, and subtract the effect of the disc and bulge. However, rotation curve measurements consist of ionized gas, \ion{H}{1}, or stellar kinematics, none of which trace the potential as precisely as we require. Stellar rotation suffers from asymmetric drift (cf. Eq.~4.228 of Binney \& Tremaine 2008), while \ion{H}{1} has dispersions of $\sim$5-25 km~s$^{-1}$ \citep{tamburroetal}. As we argue in PWC, the best method of obtaining the circular velocity may be to calculate the asymmetric drift of the stars and add that to the observed stellar rotation. Thus, we will use the circular velocity from PWC as part of our data set. Since the asymmetric drift is itself highly sensitive to several effects (in particular, it can be made very close to zero if one assumes that the velocity ellipsoid is spherically aligned), we will also test adiabatic contraction assuming that the gas rotation \emph{does} trace the gravitational potential. This scenario should be treated as a lower limit on the possible circular velocity curves. Fig.~\ref{fig:6503rotation} presents both data sets.

\subsection{Potential complications}

The effect of a bar on the rotation curve is a potential complication, as NGC 6503 is now known to possess an end-on bar \citep{freelandetal}. Because the bar is end-on, it is unlikely to significantly affect measurements of either the stellar or gas rotation, and we therefore do not anticipate any significant bias to our results. The presence of the bar is intimately connected to the question of the correct disc density profile to use. PWC found that an inner truncated disc may provide a good fit to the observations (their scenario K). %. However, inner truncations are an unnecessary complication for our purposes. Most of the scenarios suggested for their formation (see Anderson et al. 2004 and references therein) involve some sort of secular evolution based process, and therefore the conditions for adiabatic contraction do not apply. 
The close relationship between measured inner truncations and bars \citep{andersonetal} suggests that the inner truncation in NGC 6503 may be due to the bar, reducing the problem to that of fitting an exponential disc and evaluating the consistency of the resulting fits with the disc stability properties required to reproduce the observed bar. To evaluate disc stability, we determine the stability parameters $Q$ and $X$, which measure local and global stability, respectively. %The fits that employ the circular velocity from PWC allow a direct comparison with the stability results of PWC; however, the fits that employ the ionized gas data do not allow such a direct comparison since PWC did not attempt to fit an exponential disc to the ionized gas data. We can nonetheless evaluate the disc stability parameters $Q$ and $X$ to infer disc stability.

A more serious complication is the possible presence of a massive bulge. The bulge of NGC 6503, like most other late-type galaxies, is a pseudobulge, and is therefore likely to have been formed by secular evolution. In such a process, bars and spirals cause gas to lose angular momentum, funnelling the gas into the centre of the galaxy where stars form. Therefore, the conditions for adiabatic contraction are not satisfied and we restrict the analysis to cases where baryon condensation only results in an exponential disc. However, we must still account for the possibility of a massive bulge impacting the observed rotation curve -- PWC have shown that the bulge mass is $\lesssim 5\%$ of the disc mass, arguably too large to be ignored. We account for this by adding the circular velocity due to the bulge found in PWC to the model rotation used in our fits.

\subsection{Our tests}

\begin{figure}
\begin{center}
 \includegraphics[angle=270,scale=0.6]{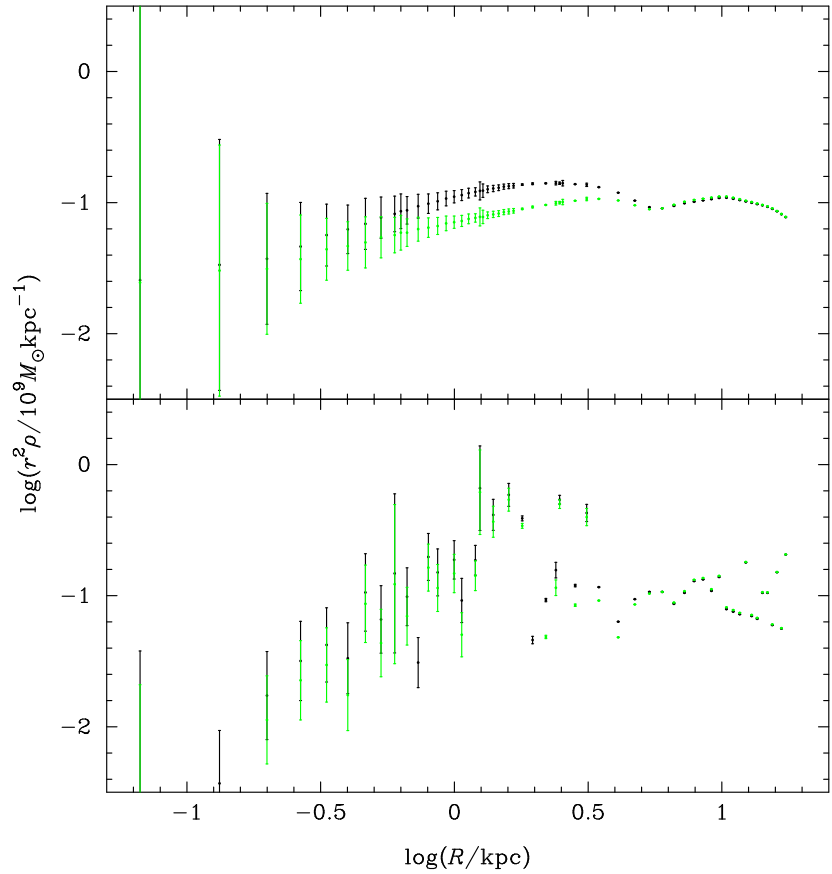}\end{center}
 \caption{The $r^2\rho$ profile for NGC 6503, estimated from the circular velocity curve (top) and ionized gas curve (bottom). The circular velocity yields double humped profiles assuming no disc (black curve) and assuming a disc of $3.5\times10^9M_{\odot}$ (green). The ionized gas curve provides noisy profiles with little useful data.}
 \label{fig:r2rho}
 \end{figure}

We first note that direct measurements of the density profile using the rotation curve are infeasible. In Fig.~\ref{fig:r2rho} we show the $r^2\rho$ profile for NGC 6503 obtained from the circular velocity and ionized gas data. The profile obtained from the circular velocity curve is shown assuming no disc and assuming a disc of $3.5\times10^9M_{\odot}$. The profile displays two peaks (owing to the kink in the circular velocity curve at $\sim7$ kpc), but there is a clear global maximum at 2.5 kpc (assuming no disc) and 10 kpc (assuming a massive disc). The central power cannot be well constrained using the available data, however. The $r^2\rho$ profile for the ionized gas data appears as little more than random noise.

We therefore adopt the following approach. We employ two separate data sets for two different suites of models: the circular velocity from PWC for which the suite of fits is labelled C, and the H$\beta$/\ion{H}{1} data from \citet{bottema89} and \citet{begeman87}, for which the suite of fits is labelled I. The baryonic components may include a bulge and disc, or solely a disc. There are thus four possible ways to conduct the fit. Each set of four runs is implemented assuming three contraction prescriptions: the adiabatic contraction prescription of \citet{blumenthaletal} which we label B, the modified adiabatic contraction prescription of \citet{gnedinetal}, labelled G, and the prescription of \citet{abadietal}, labelled A. Each of these prescriptions is applied to the standard Einasto profile. Finally, we conduct a final suite of fits assuming no baryon-induced changes to the standard Einasto profile, which we label E.

In principle, we may fit either the total rotation curve or the halo rotation obtained by subtracting the baryonic rotation from the total rotation. We have done both for each fit described above and verified that the results are consistent with each other, so for clarity we only present the results from fitting the total rotation curve.

We use a Bayesian/MCMC approach to find the best fit for each case. The choice of priors may be either uniform, or derived from the concentration-mass relation obtained by, for example, \citet{netoetal}, \citet{maccioetal}, or \citet{klypinetal}. Each of these authors finds a slightly different relation, but all find a tightly constrained concentration-mass relation of the form $C=kM_{\rm{vir}}^\epsilon$ where $k\sim10$ and $\epsilon\sim-0.1$, severely restricting the available range of parameters. In this work the $C$-$M$ relation from \citet{maccioetal} is used; because $C$ and $M_{\rm{vir}}$ depend on $\rho_{\mathrm{h}}$, $r_{\mathrm{h}}$, and $\alpha$, the $C$-$M$ relation implicitly defines priors for these parameters. We find little differences in our results for both sets of priors, so in what follows we present results for the $C$-$M$ priors only. For the remaining input parameters, we assume uniform priors. Although constraints on $r_{\mathrm{b}}$ exist from \citet{bottema89} and PWC, we allow this parameter to freely vary to determine how well an adiabatic contraction prescription can reproduce these constraints. 

Two criteria must be used to constrain the adiabatic contraction process: first, identifying which model produces the best fit; second, determining consistency with the bar formation constraints found in PWC. These constraints require that the disc not be violently unstable, but because NGC 6503 is barred, a sufficiently large disc mass is required. PWC reported a best fit disc mass of $3\pm 0.4 \times10^9M_{\odot}$ for models whose circular velocity curve matches the one employed here. Only the upper end of this range is valid because lower mass discs do not form a bar. A better proxy for disc stability is the $X$ parameter, because there is a strong correlation between the strength of the resulting bar and the $X$ value of the disc. PWC found that $\langle X\rangle\simeq2.7$ was required to produce a bar of the type found in NGC 6503. Higher $\langle X\rangle$ values failed to produce a sufficiently strong bar; as a general cutoff, we assume that $\langle X\rangle\lesssim3$ is required to form a bar. Models with low $\langle X\rangle$ are likely to be too violently unstable to properly model the galaxy, but the cutoff is less well-defined. The $Q$ value provides a better-defined lower bound on stability: It is critical that the $Q$ values of the discs we find are sufficiently large to avoid fragmentation instability, which may occur if $Q<1$, and so we also require that $Q>1$ at all radii.

\subsection{Results}

\begin{figure}
\begin{center}
 \includegraphics[angle=270,scale=0.6]{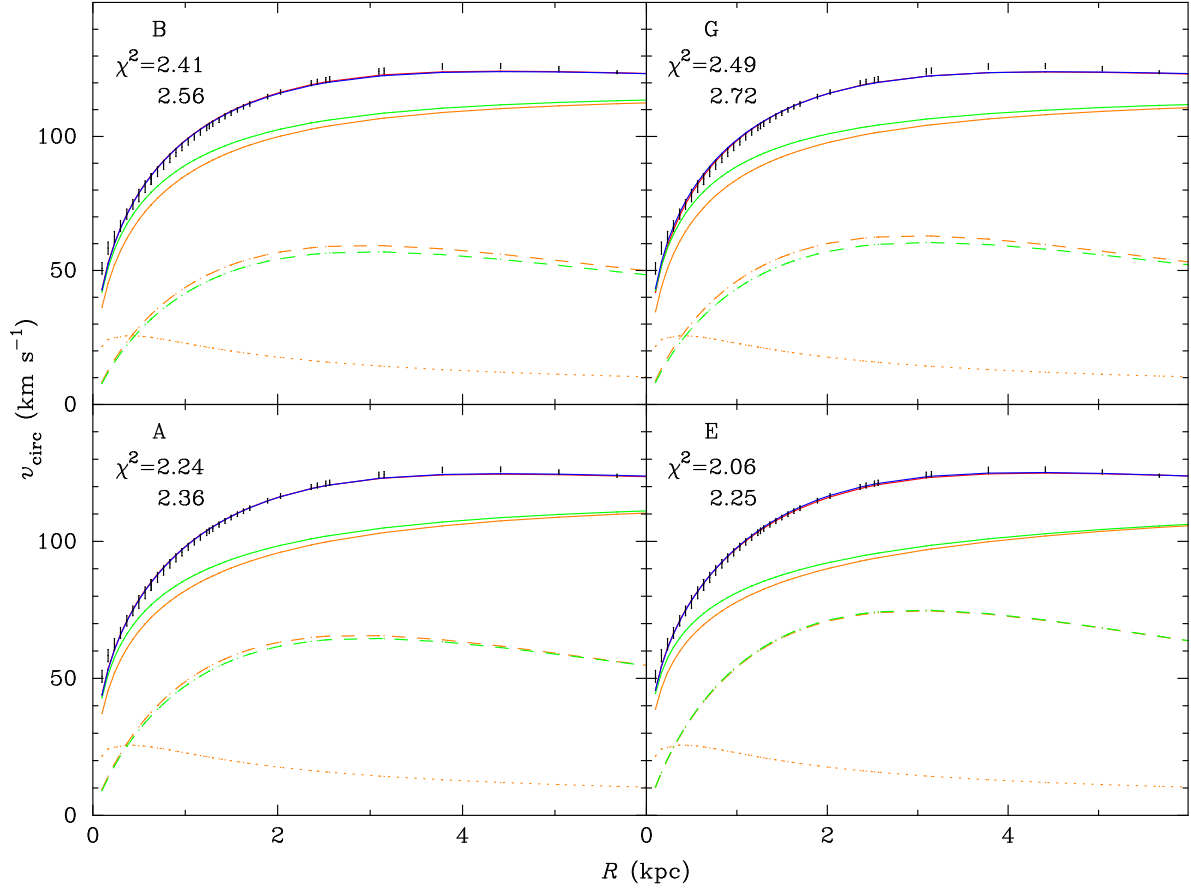}\end{center}
 \caption{Fits to the rotation curve of NGC 6503 for the circular velocity fits. Black identifies the circular velocity curve obtained from PWC, plotted at radii where the ionized gas data was obtained. Red identifies the best fit rotation curve for fits that include a bulge, and orange identifies the rotation curves of the bulge (dotted), disc (dashed) and halo (solid) components. Blue identifies the best fit for fits that include no bulge, and green identifies the rotation curves of the disc (dashed) and halo (solid) components. The best fit $\chi^2$ for each scenario is indicated, for the bulge (top number) and bulgeless (bottom number) models. The labels B, G, and A correspond to the contraction prescriptions of Blumenthal, Gnedin, and Abadi, respectively. E corresponds to an uncontracted halo.}
 \label{fig:circfit}
 \end{figure}

\begin{figure}
\begin{center}
 \includegraphics[angle=270,scale=0.6]{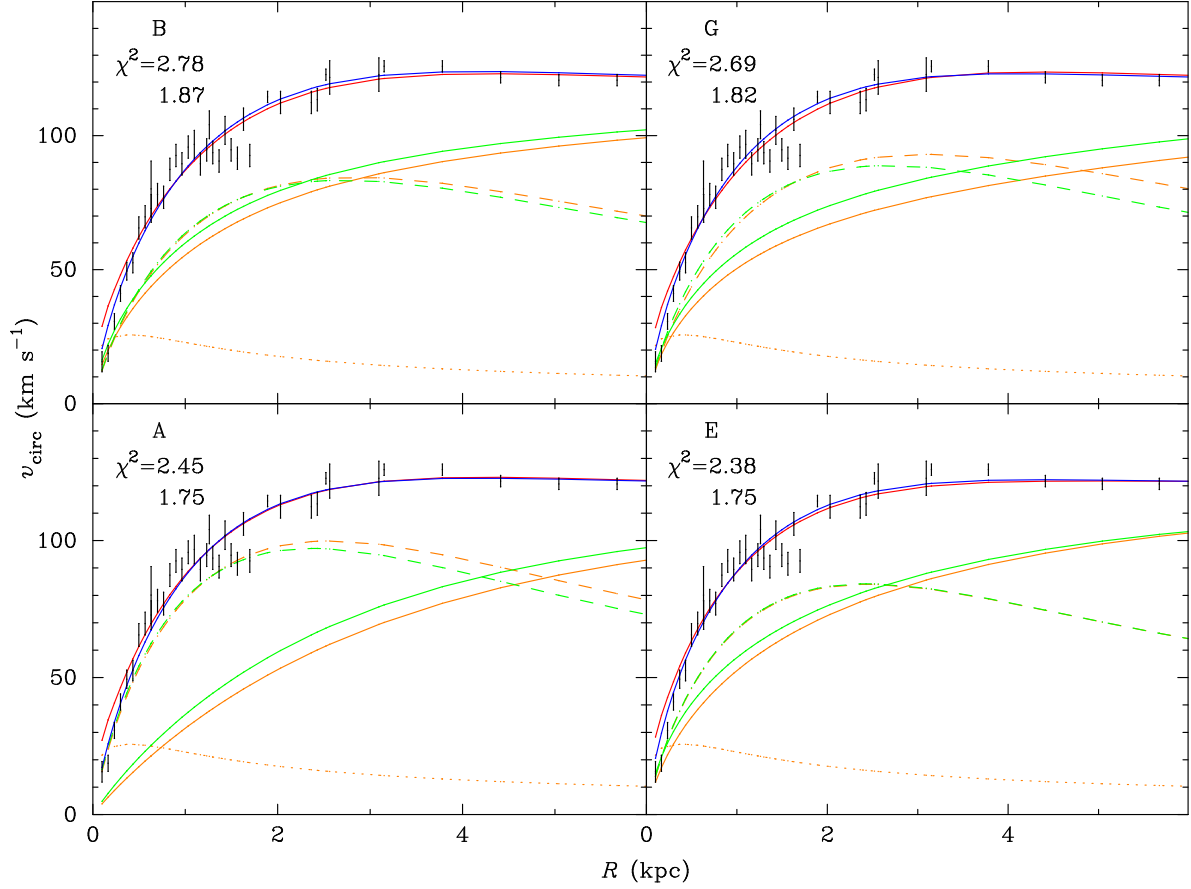}\end{center}
 \caption{Fits to the rotation curve of NGC 6503 for the ionized gas fits. Black identifies the H$\beta$/\ion{H}{1} data from \citet{bottema89} and \citet{begeman87}. Red identifies the best fit rotation curve for fits that include a bulge, and orange identifies the rotation curves of the bulge (dotted), disc (dashed) and halo (solid) components. Blue identifies the best fit for fits that include no bulge, and green identifies the rotation curves of the disc (dashed) and halo (solid) components. The best fit $\chi^2$ for each scenario is indicated, for the bulge (top number) and bulgeless (bottom number) models. The labels B, G, and A correspond to the contraction prescriptions of Blumenthal, Gnedin, and Abadi, respectively. E corresponds to an uncontracted halo.}
 \label{fig:gasfit}
 \end{figure}

Representative fits for each scenario are found in Figs.~\ref{fig:circfit} and \ref{fig:gasfit}, along with the standard $\chi^2$ statistic. The $\chi^2$ may be compared between models from suite I and between models from suite C, but should not be compared across the two types of fits because the error bars for suite C are not based on observational data. These error bars were obtained by randomly selecting several models from PWC and manually evaluating the standard deviation of the circular velocity values at each radius. The $\chi^2$ show relatively little variation, suggesting roughly equally good fits across all models; the lowest $\chi^2$ are obtained for fits to the Abadi prescription and to uncontracted halos, suggesting that the degree of contraction in NGC 6503 may be small. Interestingly, the bulgeless fits are somewhat better than the bulge fits for suite I, but there is little difference between the two for suite C.

%Several notable results present themselves.

The factor producing the largest difference between models is the rotation curve used for the fits. In particular, fits from suite I invariably produce disc masses that are much larger than the fits from suite C. The baryon fraction $f_{\mathrm{b}}$ for suite I is $\sim0.07$ and the halo masses are about $7.5\times10^{10} M_{\odot}$, producing disc masses of $\sim 6-7\times10^9 M_{\odot}$. By contrast, fits from suite C yield $f_{\mathrm{b}} \sim 0.03$ and $m_{\rm{disc}}\sim 2.5-3.5 \times 10^9 M_{\odot}$. As Table~\ref{table:results} shows, disc masses tend to increase as the effect of baryon condensation decreases, i.e., pure AC halos produce the lowest mass discs and uncontracted halos produce the heaviest discs. 

\begin{figure}
\begin{center}
 \includegraphics[angle=270,scale=0.6]{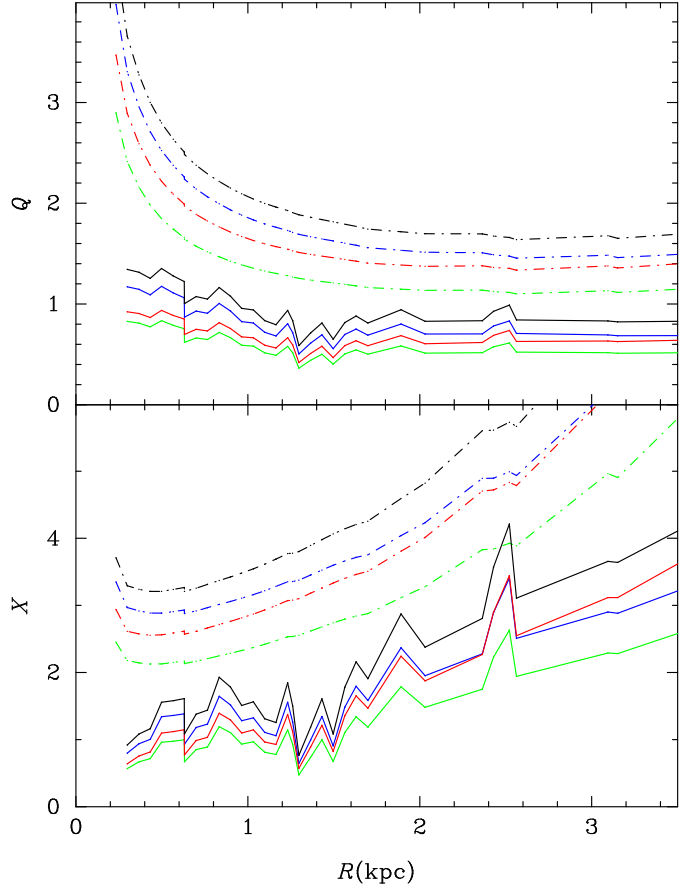}\end{center}
 \caption{Stability parameters $Q$ and $X$ for each scenario. The colour pattern black-blue-red-green (going from top to bottom for each curve) identifies scenarios B, G, A and E, respectively. Each is plotted twice, once for the circular velocity fit (dot-dashed lines) and once for the ionized gas fit (solid lines). Only the profiles for models with a bulge are shown; profiles for the bulgeless models are similar.}
 \label{fig:QandX}
 \end{figure}

To assess the bar stability of these discs, we can compare the disc masses for suite C directly to the results of PWC. The standard adiabatic contraction scenario produces discs that are too light and cannot form a bar, but the other scenarios produce discs that may be consistent with bar formation, although here too disc masses may be somewhat low. % (Fig.~\ref{fig:mdiskPDF}). 
A similar comparison is not possible for suite I, so we instead plot $Q$ and $X$ on Fig.~\ref{fig:QandX} for each fit, and present $\langle X\rangle$ in Table~\ref{table:results}. The figure shows that, except for the standard adiabatic contraction scenario, $Q$ is very close to or below 1. Below 1, the dominant instability is disc fragmentation and violent instability may result. On this basis, we reject most of the models from suite I. For the standard AC fits to the ionized gas data, the disc masses are systematically slightly smaller and hence $Q$ and $X$ are slightly larger, consistent with the $Q$ and $X$ values found in PWC that produced bars but are not violently unstable. Therefore, a scenario in which pure adiabatic contraction occurs and the circular velocity is accurately represented by the ionized gas curve is consistent with the data, but a scenario in which adiabatic contraction causes the circular velocity curve found in PWC is not consistent with the data because the discs of these models are too light (and hence $X$ too high) to form bars. The reverse holds for scenarios G, A, and E, because the disc masses in those scenarios are higher and lead to lower $Q$ and $X$. From PWC the bar formation cutoff occurs around $X_{\mathrm{min}}\sim2.2$ and therefore the circular velocity fits for scenarios A and E may be consistent with the observed bar. There is sufficient ambiguity in the derived masses and the derived $X$ and $Q$ values, however, that we are reluctant to make a definitive statement about the degree of dark matter contraction and about which models are most consistent with the data.

Thus, the overall picture that emerges suggests that stronger halo contraction correlates with lighter discs. This is because the rotation curve is fixed, and so stronger contraction means that the halo contributes more to the inner rotation curve. This trend suggests two possibilities. First, the circular velocity curve found in PWC is correct and the halo contraction is not as strong as predicted by the standard adiabatic contraction prescription. Second, the correct circular velocity curve lies between the one found in PWC and the ionized gas curve; in this case, disc masses are higher and the halo must be more strongly contracted relative to the first scenario. The two scenarios may be visualized by plotting a proxy for the degree of contraction and a proxy for disc stability. The stability proxy is simple: the best measure of disc stability is the $X$ parameter. The contraction proxy is trickier. The natural choice would be the ratio of initial to final concentrations, but the concentration is defined in terms of the virial radius, and our approach assumes that the virial radius and the outer radius for baryons $r_{\rm{outer}}$ need not coincide. Therefore, we adopt an alternate approach in which the contraction proxy is given by the ratio of halo mass inside $r_{\mathrm{b}}$ after and before contraction, $R=M_{\mathrm{f}}(r_{\mathrm{b}})/M_{\mathrm{i}}(r_{\mathrm{b}})$. 

\begin{figure}
\begin{center}
 \includegraphics[angle=270,scale=0.6]{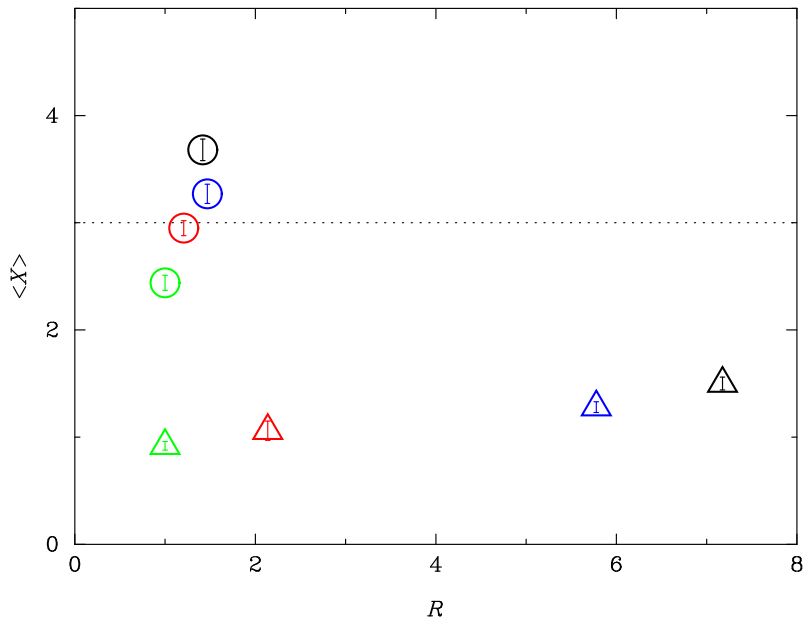}\end{center}
 \caption{Average $X$ versus contraction $R=M_{\mathrm{f}}(r_{\mathrm{b}})/M_{\mathrm{i}}(r_{\mathrm{b}})$ for each model that includes a bulge. Circles identify models from suite C and triangles identify models from suite I; the bulgeless models closely match these data points and are not plotted. Colours are as in Fig.~\ref{fig:QandX}. The dotted line indicates the approximate cutoff for bar formation from PWC; models above the cutoff are less likely to display moderate bar formation.}
 \label{fig:stabcont}
 \end{figure}

These two quantities are plotted in Fig.~\ref{fig:stabcont}, demonstrating how higher contractions are correlated with higher $\langle X\rangle$, and also reveals that most of the models are located outside the ideal stability range to produce weak bars -- in general, the fits to the circular velocity from PWC are too stable unless we assume almost no adiabatic contraction, while the fits to the ionized gas have such low $\langle X\rangle$ that they are likely to be too unstable unless we assume maximal adiabatic contraction. The ionized gas fits, represented as triangles on this figure, form a lower limit for $X$ because the true circular velocity cannot lie below the ionized gas curve. It is possible to picture suites of models that employ differing circular velocity curves; on Fig.~\ref{fig:stabcont}, these models would lie between models I and C, falling inside the ideal stability region. Therefore, the most likely scenario for NGC 6503 is that the circular velocity curve lies between the circular velocity determined in PWC and the ionized gas curve. If we instead assume that the circular velocity from PWC is correct, models that employ a standard Einasto profile without modification are most consistent with the desired stability properties. The shape parameter $\alpha$ is lower than found in the other models at $\alpha\sim 0.10$, placing it somewhat below the values obtained from theory. 

\begin{figure}
\begin{center}
 \includegraphics[angle=270,scale=0.6]{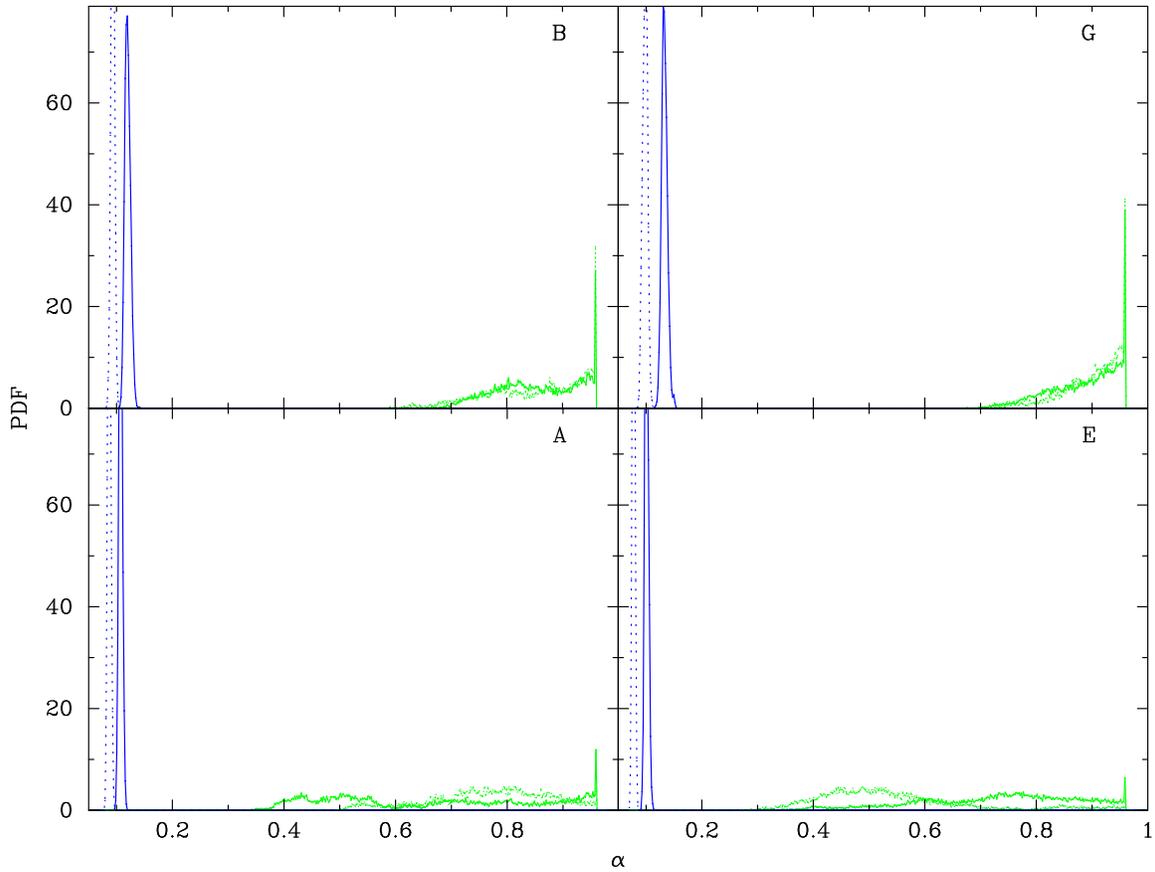}\end{center}
 \caption{PDFs for the Einasto shape parameter $\alpha$. Blue identifies the PDFs for suite C, while green identifies the PDFs for suite I. Solid lines identify the fits that include a bulge, while dotted lines identify fits that are bulgeless.}
 \label{fig:alphaPDF}
 \end{figure}

It is interesting to note that the shape parameter $\alpha$ is also severely affected by the choice of rotation curve to fit, as seen in Fig.~\ref{fig:alphaPDF}. Suite C yields $\alpha\sim0.1-0.15$, nearly consistent with (albeit slightly smaller than) those found in cosmological simulations \citep{navarroetal04, navarroetal10}. For suite I, we find $\alpha\sim 0.6-1.0$, wildly divergent from theory, and moreover that the PDFs are not well constrained. This striking difference testifies to the sensitive dependence of halo shape on the rotation curve used for the fits, and underscores the difficulty of inferring halo parameters from rotation curve observations.

\begin{figure}
\begin{center}
 \includegraphics[angle=270,scale=0.6]{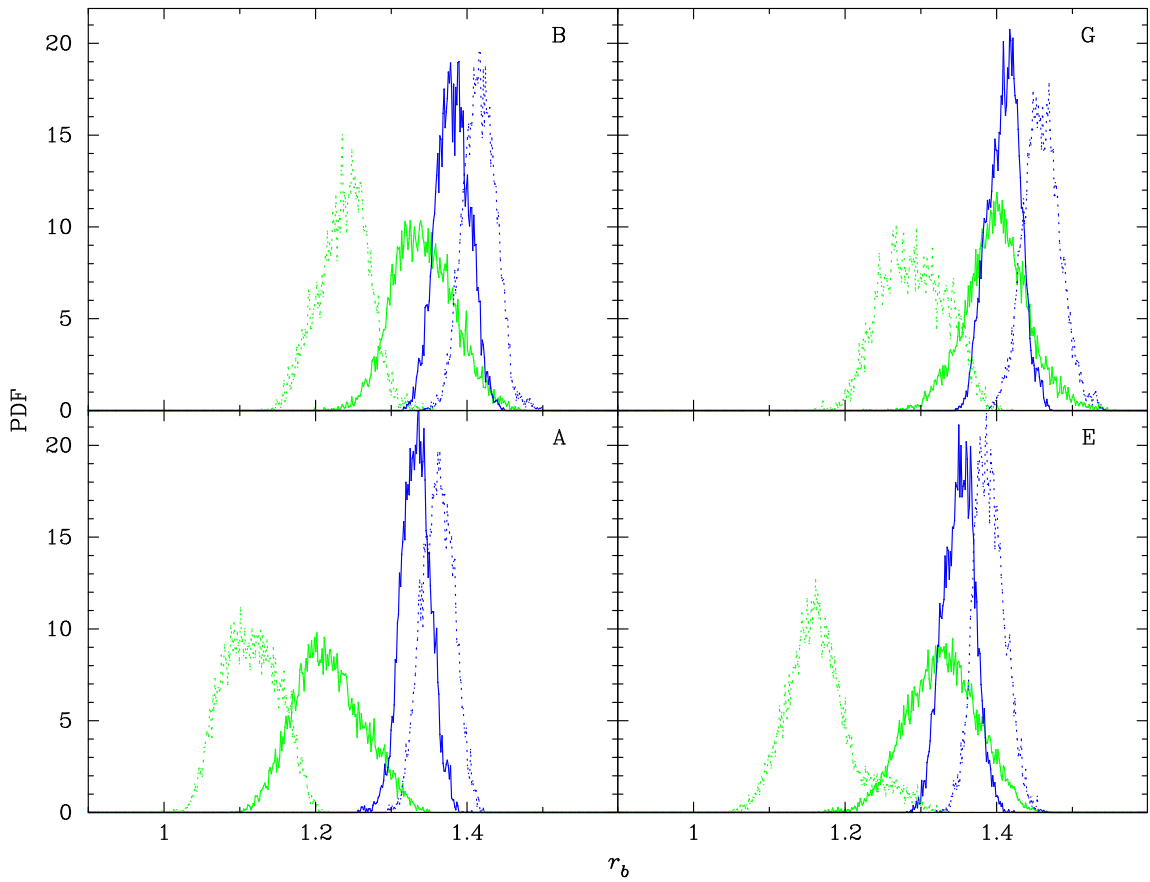}\end{center}
 \caption{PDFs for the disc scale length $r_{\mathrm{b}}$. Colours and line types are as in Fig.~\ref{fig:alphaPDF}.}
 \label{fig:rdiscPDF}
 \end{figure}

The choice of whether or not to include the bulge in the fits generally makes little difference to the overall result, because the bulge is small enough that its effect can essentially be replicated by altering the disc scale length slightly. As Fig.~\ref{fig:rdiscPDF} shows, bulgeless models for suite I produce discs that are more centrally concentrated. For suite C, the bulgeless models produce discs of slightly larger $r_{\mathrm{b}}$. Apart from this detail, however, the resulting fits are not much different and the disc scale lengths produced by our fits are consistent with the value determined in PWC of 1.3 kpc, or nearly so to within $\sim$10 \% (Table~\ref{table:results}). 

\section{Conclusion}

We have investigated the structure of adiabatically contracted Einasto profiles for the case where the baryons form an exponential disc, and obtained a number of interesting results. 

First, AC halos display power law behaviour as $r\rightarrow0$ and are well fit inside the peak of the $r^2\rho$ profile by double power laws, even in cases where the contraction is minimal. The central power index $\gamma$ is given by $0.75\pm0.18$ for Blumenthal's prescription and $0.61\pm0.08$ for Gnedin's prescription for cosmologically motivated values of the shape parameter $\alpha$. This result suggests a test for adiabatic contraction for various galaxy populations, on the assumption that uncontracted dark matter halos are well represented by Einasto profiles.

Second, attempts to fit AC Einasto profiles and their exponential discs using uncontracted halos overestimate the disc mass significantly, suggesting that adiabatic contraction must be accounted for when fitting mass models to rotation curves. However, the fits themselves are good for Einasto, NFW-type, and double power profiles. Double power and NFW-type fits using uncontracted halos tend to overestimate $\gamma$; better estimates for $\gamma$ are obtained when fitting only the inner rotation curve.

Third, we have been able to fit the circular velocity curve of NGC 6503 to three different adiabatic contraction prescriptions, and found that obtaining the correct prescription is highly dependent on the correct choice of circular velocity curve. Assuming the rotation curve found in PWC, there is almost no contraction to the halo, but lower circular velocity curves may imply stronger contraction, up to the magnitude implied by Blumenthal et al.'s adiabatic contraction prescription. Thus, the available data fails to properly constrain the degree of halo contraction, underscoring the need for accurate determination of the circular velocity curve of spiral galaxies.

\acknowledgements

The author wishes to thank L. M. Widrow for helpful discussions and critical comments on earlier versions of the manuscript, and the anonymous referee for comments and recommendations that greatly improved the manuscript.

 \clearpage
 
%  \begin{figure}
% \begin{center}
%  \includegraphics[angle=270,scale=0.6]{discmassPDF-all.ps}\end{center}
%  \caption{PDFs for the disc mass $m_{\rm{disc}}$. Blue identifies the PDFs for suite C, while green identifies the PDFs for suite I. Solid lines identify the fits that include a bulge, while dotted lines identify fits that are bulgeless.}
%  \label{fig:mdiscPDF}
%  \end{figure}
%  
%  \begin{figure}
% \begin{center}
%  \includegraphics[angle=270,scale=0.6]{baryonfracPDF-all.ps}\end{center}
%  \caption{PDFs for the baryon fraction $f$. Colours and line types are as in Fig.~\ref{fig:mdiscPDF}.}
%  \label{fig:baryonfracPDF}
%  \end{figure}
% 
 
 \clearpage

\begin{table}
\begin{center}
\begin{tabular}{lll}
 \tableline
Parameter&Units&Description\\
\tableline
$\rho_{\mathrm{h}}$&$10^9 M_{\odot}$ kpc$^{-3}$&Halo density at $r_{\mathrm{h}}$\\
$r_{\mathrm{h}}$&kpc&Radius of $r^2\rho$ peak\\
$\alpha$&Dimensionless&Einasto shape parameter\\
$f_{\mathrm{b}}$&Dimensionless&Baryon fraction\\
$r_{\mathrm{b}}$&kpc&Scale length of the final disc\\
$r_{\rm{outer}}$&kpc&Radius beyond which no baryons condense\\
\tableline
Calculated Quantities\\
\tableline
$m_{\rm{disc}}$&$10^9 M_{\odot}$&Disc mass\\
$m_{\rm{halo}}$&$10^9 M_{\odot}$&Halo mass inside $r_{\rm{outer}}$\\
$C$& &Halo concentration\\
$\langle X\rangle$& &Average $X$ over two disc scale lengths\\
\tableline
\end{tabular}
\end{center}
\caption{Table of the input parameters as well as calculated quantities that are output for each model.}
\label{table:parametertable}
\end{table}

%\renewcommand{\baselinestretch}{1.00}\normalsize
%\begin{scriptsize}
\begin{table}
\begin{center}
\renewcommand{\arraystretch}{1}
\scriptsize
\begin{tabular}{lrlrlrlrl}
\tableline
Parameter    &   \multicolumn{8}{c}{Best-fit value}  \\                                        
\tableline
\multicolumn{9}{c}{Einasto fits, Blumenthal prescription}  \\                                        
\tableline
&  \multicolumn{2}{c}{$f=0.025$}  &      \multicolumn{2}{c}{$f=0.05$}      &       \multicolumn{2}{c}{$f=0.075$}      &       \multicolumn{2}{c}{$f=0.1$}  \\                                        
\tableline
$ \rho_0 $ & $ 0.0022 $ & $ \pm 0.0002 $ & $ 0.0018 $ & $ \pm 0.0002 $ & $ 0.0018 $ & $ \pm 0.0003 $ & $ 0.0020 $ & $ \pm 0.0004 $ \\
$ r_{\mathrm{h}} $ & $ 17.6 $ & $ \pm 0.6 $ & $ 19.0 $ & $ \pm 1.3 $ & $ 18.4 $ & $ \pm 1.2 $ & $ 17.3 $ & $ \pm 1.6 $ \\
$ \alpha $ & $ 0.130 $ & $ \pm 0.003 $ & $ 0.123 $ & $ \pm 0.004 $ & $ 0.125 $ & $ \pm 0.004 $ & $ 0.128 $ & $ \pm 0.006 $ \\
$ m_{\rm{disc}} $ & $ 40.3 $ & $ \pm 1.7 $ & $ 40.7 $ & $ \pm 1.4 $ & $ 39.8 $ & $ \pm 1.1 $ & $ 38.8 $ & $ \pm 1.1 $ \\
%$ r_{\mathrm{b}} $ & $ 3 $ & $ \pm 0 $ & $ 3 $ & $ \pm 0 $ & $ 3 $ & $ \pm 0 $ & $ 3 $ & $ \pm 0 $ \\
%$ m_{\rm{halo}} $ & $ 1100 $ & $ \pm 6.88 $ & $ 1080 $ & $ \pm 27.2 $ & $ 1010 $ & $ \pm 31.8 $ & $ 931 $ & $ \pm 43.7 $ \\
%$ \chi^2 $ & $ 64.3 $ & &     $ 64.5 $ & &     $ 72.1 $ & &     $ 72.3 $ &     \\
\tableline
\multicolumn{9}{c}{Einasto fits, Gnedin prescription}  \\                                        
\tableline
&  \multicolumn{2}{c}{$f=0.025$}  &      \multicolumn{2}{c}{$f=0.05$}      &       \multicolumn{2}{c}{$f=0.075$}      &       \multicolumn{2}{c}{$f=0.1$}  \\                                        
\tableline
$ \rho_0 $ & $ 0.0023 $ & $ \pm 0.0002 $ & $ 0.0018 $ & $ \pm 0.0002 $ & $ 0.0018 $ & $ \pm 0.0003 $ & $ 0.0019 $ & $ \pm 0.0003 $ \\
$ r_{\mathrm{h}} $ & $ 17.1 $ & $ \pm 0.6 $ & $ 19.2 $ & $ \pm 1.1 $ & $ 19.0 $ & $ \pm 1.6 $ & $ 17.8 $ & $ \pm 1.5 $ \\
$ \alpha $ & $ 0.125 $ & $ \pm 0.003 $ & $ 0.118 $ & $ \pm 0.004 $ & $ 0.119 $ & $ \pm 0.005 $ & $ 0.123 $ & $ \pm 0.005 $ \\
$ m_{\rm{disc}} $ & $ 34.3 $ & $ \pm 1.6 $ & $ 37.8 $ & $ \pm 1.3 $ & $ 37.9 $ & $ \pm 1.3 $ & $ 37.4 $ & $ \pm 1.0 $ \\
%$ r_{\mathrm{b}} $ & $ 3 $ & $ \pm 0 $ & $ 3 $ & $ \pm 0 $ & $ 3 $ & $ \pm 0 $ & $ 3 $ & $ \pm 0 $ \\
%$ m_{\rm{halo}} $ & $ 1110 $ & $ \pm 6.91 $ & $ 1110 $ & $ \pm 23.4 $ & $ 1060 $ & $ \pm 40.7 $ & $ 979 $ & $ \pm 38.8 $ \\
%$ \chi^2 $ & $ 70.7 $ & &     $ 72.1 $ & &     $ 82.9 $ & &     $ 82.6 $ &    \\
\tableline
\multicolumn{9}{c}{NFW fits, Blumenthal prescription}  \\                                        
\tableline
&  \multicolumn{2}{c}{$f=0.025$}  &      \multicolumn{2}{c}{$f=0.05$}      &       \multicolumn{2}{c}{$f=0.075$}      &       \multicolumn{2}{c}{$f=0.1$}  \\                                        
\tableline
$ \rho_{\mathrm{s}} $ & $ 0.0017 $ & $ \pm 0.0002 $ & $ 0.0017 $ & $ \pm 0.0002 $ & $ 0.0018 $ & $ \pm 0.0003 $ & $ 0.0024 $ & $ \pm 0.0005 $ \\
$ r_{\mathrm{s}} $ & $ 34.6 $ & $ \pm 1.4 $ & $ 33.6 $ & $ \pm 2.0 $ & $ 31.7 $ & $ \pm 2.3 $ & $ 27.1 $ & $ \pm 2.5 $ \\
$ \alpha $ & $ 3 $ &  & $ 3 $ &  & $ 3 $ &  & $ 3 $ &  \\
$ \beta $ & $ 1 $ &  & $ 1 $ &  & $ 1 $ &  & $ 1 $ &  \\
$ \gamma $ & $ 1.35 $ & $ \pm 0.02 $ & $ 1.36 $ & $ \pm 0.01 $ & $ 1.35 $ & $ \pm 0.01 $ & $ 1.33 $ & $ \pm 0.01 $ \\
$ m_{\rm{disc}} $ & $ 49.5 $ & $ \pm 1.6 $ & $ 47.6 $ & $ \pm 1.3 $ & $ 46.3 $ & $ \pm 1.1 $ & $ 44.2 $ & $ \pm 1.1 $ \\
%$ r_{\mathrm{b}} $ & $ 3 $ & $ \pm 0 $ & $ 3 $ & $ \pm 0 $ & $ 3 $ & $ \pm 0 $ & $ 3 $ & $ \pm 0 $ \\
%$ m_{\rm{halo}} $ & $ 1080 $ & $ \pm 6.12 $ & $ 1010 $ & $ \pm 17.5 $ & $ 935 $ & $ \pm 24.1 $ & $ 831 $ & $ \pm 30 $ \\
%$ \chi^2 $ & $ 74.8 $ & &     $ 84.1 $ & &     $ 112 $ & &     $ 123 $ &   \\
\tableline
\multicolumn{9}{c}{NFW fits, Gnedin prescription}  \\                                        
\tableline
&  \multicolumn{2}{c}{$f=0.025$}  &      \multicolumn{2}{c}{$f=0.05$}      &       \multicolumn{2}{c}{$f=0.075$}      &       \multicolumn{2}{c}{$f=0.1$}  \\                                        
\tableline
$ \rho_{\mathrm{s}} $ & $ 0.0015 $ & $ \pm 0.0001 $ & $ 0.0015 $ & $ \pm 0.0002 $ & $ 0.0016 $ & $ \pm 0.0002 $ & $ 0.0032 $ & $ \pm 0.0005 $ \\
$ r_{\mathrm{s}} $ & $ 36.4 $ & $ \pm 1.4 $ & $ 35.8 $ & $ \pm 2.0 $ & $ 34.3 $ & $ \pm 2.3 $ & $ 24.0 $ & $ \pm 1.6 $ \\
$ \alpha $ & $ 3 $ &  & $ 3 $ &  & $ 3 $ &  & $ 3 $ &  \\
$ \beta $ & $ 1 $ &  & $ 1 $ &  & $ 1 $ &  & $ 1 $ &  \\
$ \gamma $ & $ 1.38 $ & $ \pm 0.02 $ & $ 1.38 $ & $ \pm 0.01 $ & $ 1.37 $ & $ \pm 0.01 $ & $ 1.33 $ & $ \pm 0.01 $ \\
$ m_{\rm{disc}} $ & $ 44.2 $ & $ \pm 1.6 $ & $ 44.9 $ & $ \pm 1.2 $ & $ 44.9 $ & $ \pm 1.0 $ & $ 41.2 $ & $ \pm 1.0 $ \\
%$ r_{\mathrm{b}} $ & $ 3 $ & $ \pm 0 $ & $ 3 $ & $ \pm 0 $ & $ 3 $ & $ \pm 0 $ & $ 3 $ & $ \pm 0 $ \\
%$ m_{\rm{halo}} $ & $ 1090 $ & $ \pm 6.2 $ & $ 1030 $ & $ \pm 17 $ & $ 973 $ & $ \pm 24.4 $ & $ 811 $ & $ \pm 21.2 $ \\
%$ \chi^2 $ & $ 86.2 $ & &     $ 102 $ & &     $ 139 $ & &     $ 152 $ &    \\
\tableline
\multicolumn{9}{c}{Double power fits, Blumenthal prescription}  \\                                        
\tableline
&  \multicolumn{2}{c}{$f=0.025$}  &      \multicolumn{2}{c}{$f=0.05$}      &       \multicolumn{2}{c}{$f=0.075$}      &       \multicolumn{2}{c}{$f=0.1$}  \\                                        
\tableline
$ \rho_{\mathrm{s}} $ & $ 0.0042 $ & $ \pm 0.0004 $ & $ 0.0037 $ & $ \pm 0.0005 $ & $ 0.0062 $ & $ \pm 0.0003 $ & $ 0.0068 $ & $ \pm 0.0003 $ \\
$ r_{\mathrm{s}} $ & $ 28.7 $ & $ \pm 3.8 $ & $ 23.5 $ & $ \pm 3.2 $ & $ 40.4 $ & $ \pm 2.3 $ & $ 30.9 $ & $ \pm 2.2 $ \\
$ \alpha $ & $ 3.01 $ & $ \pm 0.12 $ & $ 2.76 $ & $ \pm 0.13 $ & $ 3.30 $ & $ \pm 0.09 $ & $ 3.15 $ & $ \pm 0.13 $ \\
$ \beta $ & $ 0.76 $ & $ \pm 0.09 $ & $ 0.92 $ & $ \pm 0.16 $ & $ 0.52 $ & $ \pm 0.02 $ & $ 0.56 $ & $ \pm 0.03 $ \\
$ \gamma $ & $ 1.24 $ & $ \pm 0.04 $ & $ 1.31 $ & $ \pm 0.04 $ & $ 1.11 $ & $ \pm 0.02 $ & $ 1.13 $ & $ \pm 0.02 $ \\
$ m_{\rm{disc}} $ & $ 44.9 $ & $ \pm 1.7 $ & $ 46.0 $ & $ \pm 1.5 $ & $ 42.4 $ & $ \pm 1.1 $ & $ 41.5 $ & $ \pm 1.3 $ \\
%$ r_{\mathrm{b}} $ & $ 3 $ & $ \pm 0 $ & $ 3 $ & $ \pm 0 $ & $ 3 $ & $ \pm 0 $ & $ 3 $ & $ \pm 0 $ \\
%$ m_{\rm{halo}} $ & $ 1090 $ & $ \pm 7.21 $ & $ 1080 $ & $ \pm 27.6 $ & $ 1020 $ & $ \pm 42.8 $ & $ 958 $ & $ \pm 71.7 $ \\
%$ \chi^2 $ & $ 71.1 $ & &     $ 80.4 $ & &     $ 86.1 $ & &     $ 91.8 $ &   \\
\tableline
\multicolumn{9}{c}{Double power fits, Gnedin prescription}  \\                                        
\tableline
&  \multicolumn{2}{c}{$f=0.025$}  &      \multicolumn{2}{c}{$f=0.05$}      &       \multicolumn{2}{c}{$f=0.075$}      &       \multicolumn{2}{c}{$f=0.1$}  \\                                        
\tableline
$ \rho_{\mathrm{s}} $ & $ 0.0044 $ & $ \pm 0.0004 $ & $ 0.0044 $ & $ \pm 0.0002 $ & $ 0.0069 $ & $ \pm 0.0005 $ & $ 0.0068 $ & $ \pm 0.0006 $ \\
$ r_{\mathrm{s}} $ & $ 35.0 $ & $ \pm 1.4 $ & $ 39.3 $ & $ \pm 3.1 $ & $ 37.3 $ & $ \pm 1.7 $ & $ 32.3 $ & $ \pm 2.2 $ \\
$ \alpha $ & $ 3.17 $ & $ \pm 0.04 $ & $ 3.18 $ & $ \pm 0.10 $ & $ 3.22 $ & $ \pm 0.11 $ & $ 3.17 $ & $ \pm 0.12 $ \\
$ \beta $ & $ 0.64 $ & $ \pm 0.02 $ & $ 0.57 $ & $ \pm 0.03 $ & $ 0.52 $ & $ \pm 0.02 $ & $ 0.56 $ & $ \pm 0.03 $ \\
$ \gamma $ & $ 1.21 $ & $ \pm 0.02 $ & $ 1.19 $ & $ \pm 0.02 $ & $ 1.13 $ & $ \pm 0.01 $ & $ 1.14 $ & $ \pm 0.01 $ \\
$ m_{\rm{disc}} $ & $ 38.1 $ & $ \pm 1.6 $ & $ 40.5 $ & $ \pm 1.3 $ & $ 40.1 $ & $ \pm 1.5 $ & $ 39.6 $ & $ \pm 1.4 $ \\
%$ r_{\mathrm{b}} $ & $ 3 $ & $ \pm 0 $ & $ 3 $ & $ \pm 0 $ & $ 3 $ & $ \pm 0 $ & $ 3 $ & $ \pm 0 $ \\
%$ m_{\rm{halo}} $ & $ 1100 $ & $ \pm 7.04 $ & $ 1100 $ & $ \pm 32.2 $ & $ 1060 $ & $ \pm 66.4 $ & $ 980 $ & $ \pm 82.9 $ \\
%$ \chi^2 $ & $ 80.2 $ & &     $ 86.5 $ & &     $ 104 $ & &     $ 110 $ &    \\
\tableline

\end{tabular}
\end{center}
\caption{Posterior mean values for the input parameters with 1$-\sigma$ error bars for the suite of test rotation curve fits.}
\label{table:testresults}
\end{table}

{%
\begin{table}
\newcommand{\mc}[3]{\multicolumn{#1}{#2}{#3}}
\begin{center}
\begin{tabular}{lrlrlrlrl}
\tableline
\mc{9}{c}{Inner rotation fits, Blumenthal prescription}\\
\tableline
× & \mc{2}{c}{$f=0.025$} & \mc{2}{c}{$f=0.05$} & \mc{2}{c}{$f=0.075$} & \mc{2}{c}{$f=0.1$}\\
\tableline
$\beta$ & \mc{1}{r}{0.63} & $\pm 0.18$ & \mc{1}{r}{0.91} & $\pm 0.28$ & \mc{1}{r}{0.96} & $\pm 0.18$ & \mc{1}{r}{0.75} & $\pm 0.21$\\
$\gamma$ & \mc{1}{r}{0.77} & $\pm 0.09$ & \mc{1}{r}{0.92} & $\pm 0.09$ & \mc{1}{r}{0.93} & $\pm 0.07$ & \mc{1}{r}{0.80} & $\pm 0.08$\\
$\delta$ & \mc{1}{r}{3.80} & $\pm 1.07$ & \mc{1}{r}{6.92} & $\pm 1.27$ & \mc{1}{r}{6.85} & $\pm 1.27$ & \mc{1}{r}{4.03} & $\pm 0.81$\\
\tableline
\mc{9}{c}{Inner rotation fits, Gnedin prescription}\\
\tableline
× & \mc{2}{c}{$f=0.025$} & \mc{2}{c}{$f=0.05$} & \mc{2}{c}{$f=0.075$} & \mc{2}{c}{$f=0.1$}\\
\tableline
$\beta$ & \mc{1}{r}{0.50} & $\pm 0.08$ & \mc{1}{r}{1.02} & $\pm 0.57$ & \mc{1}{r}{0.79} & $\pm 0.23$ & \mc{1}{r}{0.90} & $\pm 0.19$\\
$\gamma$ & \mc{1}{r}{0.67} & $\pm 0.05$ & \mc{1}{r}{0.87} & $\pm 0.13$ & \mc{1}{r}{0.82} & $\pm 0.09$ & \mc{1}{r}{0.88} & $\pm 0.07$\\
$\delta$ & \mc{1}{r}{5.11} & $\pm 1.07$ & \mc{1}{r}{4.37} & $\pm 1.28$ & \mc{1}{r}{4.03} & $\pm 0.92$ & \mc{1}{r}{4.59} & $\pm 0.79$\\
\tableline
\end{tabular}
\end{center}
\caption{Fitted double power shape parameters to the inner ($<$5 kpc) rotation curves of the AC halos in \S2.2, with $1-\sigma$ error bars.}
\label{table:innerfits}
\end{table}
}%

{%
\begin{table}
\newcommand{\mc}[3]{\multicolumn{#1}{#2}{#3}}
\begin{center}
\begin{tabular}{lllll}
\tableline
\mc{5}{c}{Shape parameters, Blumenthal prescription}\\\cline{1-5}
\tableline
 & \mc{1}{r}{$f=0.025$} & \mc{1}{r}{$f=0.05$} & \mc{1}{r}{$f=0.075$} & \mc{1}{r}{$f=0.1$}\\
\tableline
$\beta$  & \mc{1}{r}{0.59} & \mc{1}{r}{0.53} & \mc{1}{r}{0.53} & \mc{1}{r}{0.56}\\
$\gamma$  & \mc{1}{r}{0.76} & \mc{1}{r}{0.76} & \mc{1}{r}{0.76} & \mc{1}{r}{0.77}\\
$\delta$ & \mc{1}{r}{3.23} & \mc{1}{r}{4.95} & \mc{1}{r}{4.84} & \mc{1}{r}{5.04}\\
\tableline
\mc{5}{c}{Shape parameters, Gnedin prescription}\\
\tableline
× & \mc{1}{r}{$f=0.025$} & \mc{1}{r}{$f=0.05$} & \mc{1}{r}{$f=0.075$} & \mc{1}{r}{$f=0.1$}\\
\tableline
$\beta$ & \mc{1}{r}{0.50} & \mc{1}{r}{0.55} & \mc{1}{r}{0.54} & \mc{1}{r}{0.54}\\
$\gamma$ & \mc{1}{r}{0.60} & \mc{1}{r}{0.63} & \mc{1}{r}{0.64} & \mc{1}{r}{0.64}\\
$\delta$ & \mc{1}{r}{3.40} & \mc{1}{r}{3.13} & \mc{1}{r}{3.08} & \mc{1}{r}{3.03}\\
\tableline
\end{tabular}
\end{center}
\caption{Double power shape parameters for the inner profiles of the AC halos in \S2.2, obtained directly from the $r^2\rho$ profiles.}
\label{table:truegamma}
\end{table}
}%

\begin{table}
\begin{center}
\renewcommand{\arraystretch}{1.0}
\scriptsize
\begin{tabular}{lrlrlrlrl}
\tableline
Parameter &    \multicolumn{8}{c}{Best-fit value}        \\ 
\tableline
\multicolumn{9}{c}{Suite I} \\                                       
\tableline
& \multicolumn{2}{c}{B} & \multicolumn{2}{c}{G} & \multicolumn{2}{c}{A} & \multicolumn{2}{c}{E} \\                                                                     
\tableline
$ \rho_{\rm{h}} $ & $ 0.0024 $ & $ \pm 0.0003 $ & $ 0.0025 $ & $ \pm 0.0004 $ & $ 0.0024 $ & $ \pm 0.0004 $ & $ 0.0029 $ & $ \pm 0.0004 $ \\
$ r_{\rm{h}} $ & $ 11.8 $ & $ \pm 0.9 $ & $ 11.6 $ & $ \pm 1.1 $ & $ 11.6 $ & $ \pm 0.9 $ & $ 10.1 $ & $ \pm 0.6 $ \\
$ \alpha $ & $ 0.853 $ & $ \pm 0.073 $ & $ 0.885 $ & $ \pm 0.061 $ & $ 0.664 $ & $ \pm 0.186 $ & $ 0.729 $ & $ \pm 0.142 $ \\
$ r_{\rm{b}} $ & $ 1.34 $ & $ \pm 0.04 $ & $ 1.40 $ & $ \pm 0.04 $ & $ 1.22 $ & $ \pm 0.04 $ & $ 1.3 $ & $ \pm 0.05 $ \\
$ f_{\rm{b}} $ & $ 0.058 $ & $ \pm 0.005 $ & $ 0.076 $ & $ \pm 0.005 $ & $ 0.064 $ & $ \pm 0.016 $ &  &  \\
$ r_{\rm{outer}} $ & $ 71.6 $ & $ \pm 22.3 $ & $ 72.7 $ & $ \pm 20.8 $ & $ 70.4 $ & $ \pm 19.2 $ &  &  \\
$ m_{\rm{halo}} $ & $ 100 $ & $ \pm 10 $ & $ 95 $ & $ \pm 8 $ & $ 114 $ & $ \pm 21 $ & $ 90 $ & $ \pm 18 $ \\
%$ M_{\rm{virial} $ & $ 103.0 $ & $ \pm 10.2 $ & $ 97.3 $ & $ \pm 11.9 $ & $ 122.0 $ & $ \pm 26.3 $ & $ 121.0 $ & $ \pm 10.7 $ \\
%$ R_{\rm{virial} $ & $ 93.1 $ & $ \pm 2.9 $ & $ 91.4 $ & $ \pm 2.6 $ & $ 96.9 $ & $ \pm 5.9 $ & $ 88.7 $ & $ \pm 6.8 $ \\
$ m_{\rm{disk}} $ & $ 5.8 $ & $ \pm 0.4 $ & $ 7.2 $ & $ \pm 0.5 $ & $ 7.0 $ & $ \pm 0.9 $ & $ 9.2 $ & $ \pm 0.8 $ \\
$ C $ & $ 7.91 $ & $ \pm 0.44 $ & $ 7.95 $ & $ \pm 0.51 $ & $ 8.41 $ & $ \pm 0.60 $ & $ 8.56 $ & $ \pm 0.53 $ \\
$ \langle X\rangle $ & $ 1.50 $ & $ \pm 0.06 $ & $ 1.28 $ & $ \pm 0.05 $ & $ 1.06 $ & $ \pm 0.09 $ & $ 0.92 $ & $ \pm 0.04 $ \\
\tableline
$ \chi^2 $ & $ 2.78 $  &      & $ 2.69 $  &      & $ 2.45 $ &       & $ 2.38 $      \\
\tableline
\multicolumn{9}{c}{Suite I, no bulge} \\                                       
\tableline
& \multicolumn{2}{c}{B} & \multicolumn{2}{c}{G} & \multicolumn{2}{c}{A} & \multicolumn{2}{c}{E} \\                                                                     
\tableline 
$ \rho_{\rm{h}} $ & $ 0.0029 $ & $ \pm 0.0003 $ & $ 0.0030 $ & $ \pm 0.0003 $ & $ 0.0031 $ & $ \pm 0.0003 $ & $ 0.0036 $ & $ \pm 0.00038 $ \\
$ r_{\rm{h}} $ & $ 10.8 $ & $ \pm 0.6 $ & $ 10.6 $ & $ \pm 0.5 $ & $ 10.4 $ & $ \pm 0.6 $ & $ 8.9 $ & $ \pm 0.5 $ \\
$ \alpha $ & $ 0.850 $ & $ \pm 0.086 $ & $ 0.901 $ & $ \pm 0.051 $ & $ 0.769 $ & $ \pm 0.105 $ & $ 0.541 $ & $ \pm 0.139 $ \\
$ r_{\rm{b}} $ & $ 1.24 $ & $ \pm 0.03 $ & $ 1.29 $ & $ \pm 0.04 $ & $ 1.11 $ & $ \pm 0.04 $ & $ 1.17 $ & $ \pm 0.05 $ \\
$ f_{\rm{b}} $ & $ 0.057 $ & $ \pm 0.005 $ & $ 0.077 $ & $ \pm 0.005 $ & $ 0.078 $ & $ \pm 0.011 $ &  &  \\
$ r_{\rm{outer}} $ & $ 72.6 $ & $ \pm 20.2 $ & $ 68.9 $ & $ \pm 21.6 $ & $ 47.0 $ & $ \pm 21.6 $ &  &  \\
$ m_{\rm{halo}} $ & $ 94 $ & $ \pm 8 $ & $ 87 $ & $ \pm 5 $ & $ 88 $ & $ \pm 10 $ & $ 101 $ & $ \pm 17 $ \\
%$ M_{\rm{virial} $ & $ 95.1 $ & $ \pm 7.8 $ & $ 88.8 $ & $ \pm 4.2 $ & $ 97.8 $ & $ \pm 10.1 $ & $ 133.0 $ & $ \pm 11.9 $ \\
%$ R_{\rm{virial} $ & $ 91.0 $ & $ \pm 2.4 $ & $ 89.0 $ & $ \pm 1.6 $ & $ 88.9 $ & $ \pm 3.3 $ & $ 96.9 $ & $ \pm 8.5 $ \\
$ m_{\rm{disk}} $ & $ 5.3 $ & $ \pm 0.3 $ & $ 6.7 $ & $ \pm 0.6 $ & $ 6.8 $ & $ \pm 0.6 $ & $ 7.4 $ & $ \pm 0.8 $ \\
$ C $ & $ 8.41 $ & $ \pm 0.33 $ & $ 8.39 $ & $ \pm 0.32 $ & $ 8.57 $ & $ \pm 0.57 $ & $ 10.0 $ & $ \pm 0.9 $ \\
$ \langle X\rangle $ & $ 1.42 $ & $ \pm 0.06 $ & $ 1.21 $ & $ \pm 0.05 $ & $ 0.96 $ & $ \pm 0.04 $ & $ 0.95 $ & $ \pm 0.06 $ \\
\tableline
$ \chi^2 $ & $ 1.87 $ &       & $ 1.82 $  &      & $ 1.75 $  &      & $ 1.75 $      \\
\tableline
\multicolumn{9}{c}{Suite C} \\                                       
\tableline
& \multicolumn{2}{c}{B} & \multicolumn{2}{c}{G} & \multicolumn{2}{c}{A} & \multicolumn{2}{c}{E} \\                                                                     
\tableline
$ \rho_{\rm{h}} $ & $ 0.0028 $ & $ \pm 0.0001 $ & $ 0.0028 $ & $ \pm 0.0001 $ & $ 0.0028 $ & $ \pm 0.0001 $ & $ 0.0036 $ & $ \pm 0.0002 $ \\
$ r_{\rm{h}} $ & $ 9.9 $ & $ \pm 0.2 $ & $ 9.7 $ & $ \pm 0.2 $ & $ 9.7 $ & $ \pm 0.1 $ & $ 8.5 $ & $ \pm 0.2 $ \\
$ \alpha $ & $ 0.120 $ & $ \pm 0.005 $ & $ 0.133 $ & $ \pm 0.005 $ & $ 0.107 $ & $ \pm 0.003 $ & $ 0.101 $ & $ \pm 0.003 $ \\
$ r_{\rm{b}} $ & $ 1.38 $ & $ \pm 0.02 $ & $ 1.41 $ & $ \pm 0.02 $ & $ 1.33 $ & $ \pm 0.02 $ & $ 1.35 $ & $ \pm 0.02 $ \\
$ f_{\rm{b}} $ & $ 0.026 $ & $ \pm 0.003 $ & $ 0.029 $ & $ \pm 0.002 $ & $ 0.026 $ & $ \pm 0.002 $ &  &  \\
$ r_{\rm{outer}} $ & $ 38.8 $ & $ \pm 5.7 $ & $ 41.1 $ & $ \pm 5.1 $ & $ 45.5 $ & $ \pm 4.3 $ &  &  \\
$ m_{\rm{halo}} $ & $ 117 $ & $ \pm 14 $ & $ 121 $ & $ \pm 12 $ & $ 134 $ & $ \pm 10 $ & $ 214 $ & $ \pm 11 $ \\
%$ M_{\rm{virial} $ & $ 243.0 $ & $ \pm 6.1 $ & $ 234.0 $ & $ \pm 6.3 $ & $ 255.0 $ & $ \pm 4.0 $ & $ 255.0 $ & $ \pm 14.9 $ \\
%$ R_{\rm{virial} $ & $ 97.9 $ & $ \pm 4.0 $ & $ 99.0 $ & $ \pm 3.4 $ & $ 103.0 $ & $ \pm 2.6 $ & $ 101.0 $ & $ \pm 8.5 $ \\
$ m_{\rm{disk}} $ & $ 3.0 $ & $ \pm 0.1 $ & $ 3.4 $ & $ \pm 0.1 $ & $ 3.5 $ & $ \pm 0.1 $ & $ 4.3 $ & $ \pm 0.1 $ \\
$ C $ & $ 9.89 $ & $ \pm 0.41 $ & $ 10.20 $ & $ \pm 0.28 $ & $ 10.60 $ & $ \pm 0.32 $ & $ 10.20 $ & $ \pm 0.91 $ \\
$ \langle X\rangle $ & $ 3.68 $ & $ \pm 0.10 $ & $ 3.27 $ & $ \pm 0.09 $ & $ 2.95 $ & $ \pm 0.07 $ & $ 2.44 $ & $ \pm 0.07 $ \\
\tableline
$ \chi^2 $ & $ 2.41 $ &       & $ 2.49 $  &      & $ 2.24 $ &       & $ 2.06 $      \\
\tableline
\multicolumn{9}{c}{Suite C, no bulge} \\                                       
\tableline
& \multicolumn{2}{c}{B} & \multicolumn{2}{c}{G} & \multicolumn{2}{c}{A} & \multicolumn{2}{c}{E} \\ 
\tableline
$ \rho_{\rm{h}} $ & $ 0.0027 $ & $ \pm 0.0002 $ & $ 0.0024 $ & $ \pm 0.0002 $ & $ 0.0028 $ & $ \pm 0.0001 $ & $ 0.0027 $ & $ \pm 0.0002 $ \\
$ r_{\rm{h}} $ & $ 9.9 $ & $ \pm 0.3 $ & $ 10.6 $ & $ \pm 0.5 $ & $ 9.8 $ & $ \pm 0.2 $ & $ 9.7 $ & $ \pm 0.3 $ \\
$ \alpha $ & $ 0.093 $ & $ \pm 0.003 $ & $ 0.098 $ & $ \pm 0.004 $ & $ 0.087 $ & $ \pm 0.003 $ & $ 0.078 $ & $ \pm 0.002 $ \\
$ r_{\rm{b}} $ & $ 1.42 $ & $ \pm 0.02 $ & $ 1.46 $ & $ \pm 0.02 $ & $ 1.36 $ & $ \pm 0.02 $ & $ 1.39 $ & $ \pm 0.02 $ \\
$ f_{\rm{b}} $ & $ 0.024 $ & $ \pm 0.003 $ & $ 0.032 $ & $ \pm 0.005 $ & $ 0.027 $ & $ \pm 0.002 $ &  &  \\
$ r_{\rm{outer}} $ & $ 37.4 $ & $ \pm 4.2 $ & $ 33.6 $ & $ \pm 5.3 $ & $ 40.2 $ & $ \pm 3.6 $ &  &  \\
$ m_{\rm{halo}} $ & $ 115 $ & $ \pm 11 $ & $ 104 $ & $ \pm 14 $ & $ 123 $ & $ \pm 9 $ & $ 242 $ & $ \pm 19 $ \\
%$ M_{\rm{virial} $ & $ 257.0 $ & $ \pm 3.6 $ & $ 251.0 $ & $ \pm 4.7 $ & $ 264.0 $ & $ \pm 3.4 $ & $ 244.0 $ & $ \pm 14.0 $ \\
%$ R_{\rm{virial} $ & $ 97.4 $ & $ \pm 3.1 $ & $ 94.1 $ & $ \pm 4.4 $ & $ 99.7 $ & $ \pm 2.5 $ & $ 87.9 $ & $ \pm 6.0 $ \\
$ m_{\rm{disk}} $ & $ 2.8 $ & $ \pm 0.1 $ & $ 3.2 $ & $ \pm 0.1 $ & $ 3.3 $ & $ \pm 0.1 $ & $ 4.4 $ & $ \pm 0.1 $ \\
$ C $ & $ 9.83 $ & $ \pm 0.54 $ & $ 8.93 $ & $ \pm 0.73 $ & $ 10.20 $ & $ \pm 0.46 $ & $ 9.39 $ & $ \pm 1.25 $ \\
$ \langle X\rangle $ & $ 4.04 $ & $ \pm 0.10 $ & $ 3.57 $ & $ \pm 0.09 $ & $ 3.20 $ & $ \pm 0.08 $ & $ 2.51 $ & $ \pm 0.06 $ \\
\tableline
$ \chi^2 $ & $ 2.56 $  &      & $ 2.72 $  &      & $ 2.36 $  &      & $ 2.25 $      \\
\tableline
\end{tabular}
\end{center}
\caption{Table of the best fit parameter values for NGC 6503 rotation curve fits with 1$-\sigma$ error bars.}
\label{table:results}
\end{table}
%\end{scriptsize}
\end{document}